\documentclass[11pt]{article}
\usepackage{graphicx,texdraw,color}
\def\today{\ifcase \month\or
  January\or February\or March\or April\or May\or June\or
    July\or August\or September\or October\or November\or December\fi
      \space\number\day,\space
        \number\year }
\newcommand{\be}{\begin{equation}}
\newcommand{\en}{\end{equation}}
\newcommand{\bea}{\begin{eqnarray}}
\newcommand{\ena}{\end{eqnarray}}
\newcommand{\lbl}[1]{\label{eq:#1}}
\newcommand{\rf}[1]{(\ref{eq:#1})}
\newcommand{\lbltab}[1]{\label{tab:#1}}

\newcommand{\qa}{   {\textcolor{red}{\Large $\bullet$} } }
\newcommand{\qv}{   {\textcolor{red}{\Large $\circ$} } }
\newcommand{\vmu}{   {\textcolor{green}{\Large x} } }
\newcommand{\amu}{   {\textcolor{blue}{\Large +} } }
\newcommand{\mad}{M^2_A}
\newcommand{\mvd}{M^2_V}
\newcommand{\msd}{M^2_S}
\newcommand{\mpid}{{M^2_\pi}}
\newcommand{\mkd}{{M^2_K}}
\newcommand{\fpi}{F_\pi}
\newcommand{\fpid}{F^2_\pi}
\newcommand{\mbd}{M^2_B}
\newcommand{\fad}{{F_A^2}}
\newcommand{\fvd}{{F_V^2}}
\newcommand{\gvfv}{{G_VF_V}}
\newcommand{\gvd}{{G^2_V}}
\newcommand{\mg}{M_\gamma}
\newcommand{\mgd}{M^2_\gamma}
\newcommand{\lngmu}{\log{\mgd\over\mu^2}}
\newcommand{\logA}{\log{\mad\over\Lambda^2}}
\newcommand{\logV}{\log{\mvd\over\Lambda^2}}
\newcommand{\logS}{\log{\msd\over\Lambda^2}}

\newcommand{\loglb}{{\log\Lambda^2\over16\pi^2}}

\newcommand{\logBV}{\log{\mbd\over\mvd}}
\newcommand{\logz}{\log{z}}
\newcommand{\logg}{\log{\mgd\over\mvd}}
\newcommand{\eone}{\mathbf{e_1}}
\newcommand{\etwo}{\mathbf{e_2}}
\newcommand{\ethree}{\mathbf{e_3}}
\newcommand{\efour}{\mathbf{e_4}}
\newcommand{\undemi}{{1\over2}}
\newcommand{\trace}[1]{\langle #1 \rangle}
\newcommand{\correl}[1]{\ \mbox{$< #1 >$}}

\newcommand{\iddp}{ {-i\,d^d p\over(2\pi)^d }\,}

\newcommand{\eeps}{\epsilon^{\mu\nu\alpha\beta} }
\newcommand{\fpab}{f_{+\alpha\beta}}
\newcommand{\um}{u_\mu}
\def\Udag{{U^\dagger}}
\def\qr{{q_R}}
\def\ql{{q_L}}
\def\lag{{\cal L}}
\newcommand{\lapprox}{%
\mathrel{%
\setbox0=\hbox{$<$}\raise0.6ex\copy0\kern-\wd0\lower0.65ex\hbox{$\sim$}}}
\newcommand{\gapprox}{%
\mathrel{%
\setbox0=\hbox{$>$}\raise0.6ex\copy0\kern-\wd0\lower0.65ex\hbox{$\sim$}}}
\newcommand{\dslash}{%
\mathrel{\setbox0=\hbox{$/$}\copy0\kern-\wd0\hbox{$\partial$}}}
\newcommand{\Gslash}{%
\mathrel{\setbox0=\hbox{$/$}\copy0\kern-1.2\wd0\hbox{$G$}}}

\begin{document}
\begin{flushright}
\today \\
IISc--CHEP-5/04\\
IPNO--DR--04-05
\end{flushright}
\begin{center}

{\Large\bf Four-point correlator constraints on electromagnetic\\
chiral parameters and resonance effective Lagrangians}\footnote{Supported 
in part by IFCPAR contract 2504-1, 
by the European Union TMR
network, Contract No. HPRN-CT-2002-00311  (EURIDICE)  and by
the Department of Science and Technology, Government of India.}
\vskip 1cm

{\Large B. Ananthanarayan$^a$  and B. Moussallam$^b$}\\
\vskip 0.2cm

{\sl$^a$\  Centre for High Energy Physics}\\
{\sl Indian Institute of Science, }
{\sl Bangalore 560 012, India}
\vskip 0.3cm

{\sl$^b$\ Groupe de Physique Th\'eorique, IPN }\\
{\sl Universit\'e Paris-Sud, }
{\sl F-91406 Orsay C\'edex, France}

\vskip 2.5truecm
{\Large\bf Abstract}\\
\vskip 0.3cm

\begin{minipage}{13cm}
We pursue the analysis of a set of generalized DGMLY sum rules for
the electromagnetic chiral parameters at order $e^2p^2$ 
and discuss implications
for  effective Lagrangians with resonances.
We exploit a formalism in which charge spurions are
introduced and treated as sources. We show that no inconsistency arises
from anomalies up to quadratic order in the spurions.
We focus on the sum rules
associated with QCD 4-point  correlators which were not analyzed in detail
before. 
Convergence properties of the sum rules are deduced from a general analysis of
the form of the counterterms in the presence of electromagnetic spurions.
Following the approach in which vector and axial-vector resonances
are described with antisymmetric tensor
fields and have a chiral order, we show that the convergence constraints
are violated at chiral order four and 
can be satisfied by introducing a set of terms of  order six. The
relevant couplings get completely and uniquely 
determined from a set of generalized Weinberg sum-rule
relations. An update on the corrections to Dashen's low-energy theorem is
given. 
\end{minipage}
\end{center}

\newpage
\tableofcontents

\section{Motivation}
Virtual photons play an important role in low energy QCD for isospin breaking
phenomena and must also be taken into account in connection with 
precision experiments.
The formalism for extending the framework of the standard 
ChPT\cite{weinberg,gl84,gl85}
to include virtual photons at one-loop was developed, some time ago, 
by Urech\cite{urech}. Several 
extensions were later performed to accommodate leptons\cite{knecht-lept},
to the  weak non-leptonic sector\cite{ecker-nlw}, to the anomalous 
sector\cite{am} as well as to the baryon sector\cite{meissner-bar}.
The problem of the determination of the new set of low-energy 
coupling constants (LEC's) which appear in Urech's Lagrangian
is of obvious practical importance in relation with radiative correction
calculations. Many such calculations have been performed recently:  
low energy $\pi\pi$ or $\pi K$ scattering 
\cite{meissner-pipi,knecht-urech,kubis-meissner1,kubis-meissner2,nehme-talav,knecht-nehme,nehmemore} 
pionic as well as pi-kaonic atoms \cite{hagop,gasser,julia}, the set of
semi-leptonic decays of the Kaon\cite{cirigl-knecht,nehme}, processes
from the anomalous sector\cite{amettler-knecht,am}, the problem of 
computing the hadronic contribution of the muon $g-2$ from $\tau$ 
decay data\cite{cirigl-ecker}, the theoretical prediction of the CP-violation
ratio $\epsilon'/\epsilon$ for the 
Kaon\cite{cirigl-dono,cirigl-enp,cirigl-pen}
to mention just a few illustrative examples.
Another important application of the electromagnetic  low-energy
couplings (LEC ) 
is to the question of the chiral corrections
to Dashen's low-energy theorem\cite{dashen} 
and the determination of the light quark masses (e.g. \cite{leutwyler}). 
Interesting questions were raised recently concerning the proper definition of
the quark masses in the presence of QED\cite{gasser-rus}.

The problem of the determination of the EM LEC's 
was addressed in several papers some time 
ago\cite{bu,bij-prad,bmdash} but the results remain incomplete and  
sometimes contradictory. In this paper, we continue the discussion started in
ref.\cite{bmdash} (hereafter referred to as (I)). In (I) it was
shown that the electromagnetic LEC's obey integral sum rule representations
which generalize the classic DGMLY sum rule\cite{dgmly} 
for the $\pi^+-\pi^0$ mass difference. 
The integral representations have the form of convolutions involving a
pure QCD $n$-point Green's function with $n=2,3$ and 4 together with the free
photon propagator. 
These representations  serve a number of purposes. For instance, 
they can be used to recover the chiral scale dependence of the LEC's, 
to determine how these depend on the $\xi$ gauge parameter and, also, by
studying the convergence properties, to determine which LEC's are affected
by short distance ambiguities. Finally, these representations can be used as
a method to provide approximate determinations of these couplings.  

This was exploited in (I) to discuss the set of couplings which are associated
with 2- and 3-point Green's functions. The method consists in 
constructing rational approximants 
to the Green's functions, which is in the spirit of the
large $N_c$ expansion\cite{thooft}. The parameters of the rational functions
are constrained by the chiral Ward identities, by physical input 
on resonance masses and couplings and by short-distance matching
conditions. In the present paper, we  address the more complicated  case
of the 4-point functions. The way of insuring  that the 
Ward identities are satisfied is to start from a chiral Lagrangian including
resonances. Such a Lagrangian was constructed in ref.\cite{egpr} which
uses the antisymmetric tensor formalism\cite{gl84} to describe 
the vector and axial-vector resonances. This formalism is particularly 
convenient for discussing the ordinary LEC's because one can assign a chiral
order to each resonance field such that all the terms which are of
order four,  which include one multiplet of
vector, axial-vector, scalar and  pseudo-scalar resonances
were listed in ref.\cite{egpr}.  
It was shown in a subsequent paper\cite{eglpr} that if one imposes 
a number of short distance matching conditions, the determination of the
$O(p^4)$ LEC's becomes independent of the specific representation of
the resonances in the chiral Lagrangian. 

The Lagrangian of ref.\cite{egpr}
was employed by Baur and Urech\cite{bu} (BU) to discuss the 
electromagnetic LEC's.
For this application, however, one realizes 
by using the integral representation that
the Green's functions are needed under kinematical conditions which 
are different, in general, from those considered in\cite{eglpr} such
that  the Lagrangian of ref.\cite{egpr} is no longer sufficient to ensure
the proper short distance constraints. 
This deficiency manifests itself
in the fact that the results of BU for the electromagnetic LEC's fail
to satisfy the correct scale dependence, which is an effect of leading order
in $N_c$. 
In this paper, we will show that a simple extension of the
Lagrangian\cite{egpr} to include a set of terms which are of chiral order six
allows one to satisfy the short distance constraints in the chiral limit 
which are necessary for a consistent determination of the electromagnetic 
LEC's and which are associated with the 2- and 3-point functions considered
in \cite{bmdash} as well as the 4-point functions introduced in this paper.
Finally, we will present estimates for the six LEC's which are associated 
with these 4-point functions. 

\section{Chiral calculation of the basic correlators}
\subsection{Definition of the  basic correlation functions}
In the presence of a dynamical photon field $F_\mu$ it is convenient to 
extend the source part of the QCD Lagrangian to include two spurion fields
$q_V$, $q_A$ \cite{urech} 
\be\lbl{lagsrc}
{\cal L}_{QCD}^{sources}= \bar\psi_L \gamma^\mu 
[\, l_\mu + q_L\, F_\mu] \psi_L 
+\bar\psi_R \gamma^\mu 
[\, r_\mu + q_R\, F_\mu] \psi_R
-\bar\psi (s-i\gamma^5 p )\psi\ ,
\en
with
\be\lbl{sources}
l_\mu=v_\mu-a_\mu,\ q_L= q_V-q_A,\ 
r_\mu=v_\mu+a_\mu,\ q_R= q_V+q_A\ .
\en
The number of light flavours is assumed to be $N^0_F=3$, 
and the sources in eq.\rf{sources} are traceless $3\times3$ matrices.
Originally\cite{urech}, the spurions $q_V,\ q_A$ were introduced only for the
purpose of classifying the independent terms in the chiral Lagrangian
and afterwards one would set $q_A=0$, $q_V=Q$. 
We will exploit here  more general applications by considering the spurions
on the same footing as the ordinary external sources. 
We note that renormalization will generate terms which are
non-linear in the sources. Those which are quadratic in the spurion
sources are of  particular interest for our purposes and will be discussed in 
detail in sec.\ref{sec:countrms}. It will also be argued that no difficulty
arises in the definition of the generating functional in the presence of
the $q_V$, $q_A$ spurions up to quadratic order in the spurions.
We now introduce a set of five correlation functions defined by taking 
two functional derivatives with respect to $v_\mu$ or $a_\mu$ and
two  derivatives with respect to the charge spurions $q_V$ or 
$q_A$. The first one  is defined as follows
\be\lbl{basedef}
1) <A_\alpha^a A_\beta^b Q^c_A Q^d_A>=
\int d^4x\, d^4y\, d^4z \exp(iky) 
{\delta^4\,W(v_\mu,a_\mu,q_V,q_A) \over
\delta a^a_\alpha(x)
\delta a^b_\beta(y)
\delta q_A^c(z)
\delta q_A^d(0) } 
\en
where $W$ is the generating functional of connected Green's functions. 
The remaining four
\bea\lbl{basedef1}
2)  <A_\alpha^a A_\beta^b Q^c_V Q^d_V>
\nonumber\\
3)  <V_\alpha^a V_\beta^b Q^c_A Q^d_A>
\nonumber\\
4)  <V_\alpha^a V_\beta^b Q^c_V Q^d_V> 
\nonumber\\
5)  <A_\alpha^a V_\beta^b Q^c_A Q^d_V> 
\ena
are defined in an exactly analogous way. 
These correlators involve two Lorentz
structures $g_{\alpha\beta}$ and $k_\alpha k_\beta$. 
For our purposes, we can restrict ourselves to the Lorentz
structure proportional to $g_{\alpha\beta}$ and we will consider the limit
$k\to 0$. Before taking this limit, however, 
we must consider two possible kinds of
singularities: a) $1/k^2$ pion poles and b) $\log(k^2)$ cuts due to pion 
loops. The latter singularities drop out in the leading large $N_c$ 
approximation, which we will frequently invoke. 
A pion pole can appear only in
one  flavour structure of the fifth correlator
$<A_\alpha^a V_\beta^b Q_A^c Q_V^d>$.
Apart from this particular case, we can simply set $k=0$, meaning that we
restrict ourselves to constant vector and axial--vector sources.
This leads to great
simplification since all terms like $\partial_\mu v_\nu$,  
$\partial_\mu a_\nu$  etc... can be dropped in the chiral resonance 
Lagrangian. Similarly, the $q_V$ and $q_A$  sources can be taken as constant
for the present applications.
 
From the definition, eq.\rf{basedef} and the explicit expression of the terms
linear in the sources eq.\rf{lagsrc} we can easily see that the correlators
introduced above can be represented as a convolution of
an ordinary connected 
QCD 4-point Green's function and the photon propagator. For 
example, one finds
\be\lbl{convol}
\correl{A_\alpha^a A_\beta^b Q^c_A Q^d_A}=\int d^4x d^4y d^4z \exp(iky) 
<0\vert T( A^a_\alpha(x) A^b_\beta(y) A^c_\rho(z) A^d_\sigma(0))\vert 0>
D_F^{\rho\sigma}(z)\ .
\en
Because of the short distance singularities, the integral over $d^4z$ is
expected to be divergent. The divergence is removed upon including the 
additional contributions stemming
from the counterterms and one may use
d-dimensional integration as a regularization method. 

Let us now examine the flavour structure. The correlators depend on four 
flavour indices.
At order $e^2p^2$ one can show by using the chiral Lagrangian that there
are only four independent tensor structures.
It will prove convenient to use the following basis 
\bea
&&\eone=f^{tac} f^{tbd}+ f^{tad} f^{tbc}
\nonumber\\
&&\etwo=\delta^{ac}\delta^{bd}+\delta^{ad}\delta^{bc}
\nonumber\\
&&\ethree=d^{tac} d^{tbd}+ d^{tad} d^{tbc}
\nonumber\\
&&\efour=f^{tab}f^{tcd}\ .
\ena
Other tensor structures that arise in practical 
calculations can be expressed in terms of this basis
by utilizing  $SU(3)$ trace identities. The following relations
are useful
\bea\lbl{flavrel}
&& d^{tab}d^{tcd} ={1\over6}{\mathbf e_1}+{1\over3}{\mathbf e_2}-{1\over2}
{\mathbf e_3}\nonumber\\
&& \delta^{ab}\delta^{cd}={1\over 2}{\mathbf e_1}
\phantom{+{1\over3}{ e_2} } +{3\over2}{\mathbf e_3}\ .
\ena
We can now expand the correlators over the flavour basis. The first
four correlators are expanded over $\eone$, $\etwo$, $\ethree$ because
of the permutation symmetry $a\leftrightarrow b$, $c\leftrightarrow d$. 
For instance, one has
\be
\correl{A_\alpha^a A_\beta^b Q^c_A Q^d_A}=ig_{\alpha\beta}\sum_{j=1}^3
\correl{AAQ_AQ_A}_j
\mathbf{e_j}\ .
\en
The expansion of the last correlator involves $\efour$ also,
\be
\correl{A_\alpha^a V_\beta^b Q^c_A Q^d_V}=ig_{\alpha\beta}\sum_{j=1}^4 
\correl{AVQ_AQ_V}_j
\mathbf{e_j}\ .
\en
This expansion apparently
generates sixteen independent flavour components. As we will see below 
there are only ten components which appear at the chiral order $e^2p^2$ 
considered here.

\subsection{Correlators  at order $e^2p^2$ }
We start by computing the correlators defined above\rf{basedef}\rf{basedef1}
in the chiral expansion up to the  order $e^2p^2$. 
Setting the sources $s$ and $p$
equal to zero, the leading order Lagrangian collects the terms of order
$e^2$ and $p^2$
\bea\lbl{lagchir2}
&&{\cal L}^{(2)}_{chir}= {\cal L}^{p^2}_{chir} + {\cal L}^{e^2}_{chir}
\nonumber\\
&&
{\cal L}^{p^2}_{chir}=
{F_0^2\over4}\trace{ D_\mu U D^\mu U^\dagger}
-{1\over4}F_{\mu\nu} F^{\mu\nu}
-{1\over 2\xi} (\partial_\mu F^\mu)^2+{1\over2}\mgd F_\mu F^\mu\ 
\nonumber\\
&& {\cal L}^{e^2}_{chir}=C\trace{ q_L U^\dagger q_R U}\ ,
\ena
with
\be
D_\mu U=\partial_\mu U-i\left( r_\mu+ q_R F_\mu \right) U
+iU \left( l_\mu+q_L F_\mu \right)\ .
\en
The photon is endowed with a small mass (which may be counted as $O(p)$) in
order to regulate infrared divergences which can appear in the chiral limit.
We will also need the following part of the chiral Lagrangian  
of order $e^2p^2$\cite{urech}
\bea
{\cal L}^{e^2p^2}_{chir}
&&={1\over2}K_1 F_0^2\trace{D_\mu U D^\mu U^\dagger}\trace{\ql\ql+\qr\qr}\\ \nonumber
&&+K_2 F_0^2\trace{D_\mu U D^\mu U^\dagger}\trace{\ql U^\dagger \qr U}\\ \nonumber
&&+K_3 F_0^2(\trace{D_\mu U\ql U^\dagger}^2 +\trace{D_\mu U^\dagger\qr U}^2)\\ \nonumber
&&+K_4 F_0^2 \trace{D_\mu U\ql U^\dagger}\trace{D_\mu U^\dagger\qr U}\\ \nonumber
&&+K_5 F_0^2\trace{D_\mu U^\dagger D^\mu U\ql\ql+ D_\mu U D^\mu U^\dagger\qr\qr}\\
\nonumber
&&+K_6 F_0^2\trace{D_\mu U^\dagger D^\mu U\ql U^\dagger \qr U +D_\mu U
D^\mu U^\dagger\qr U \ql U^\dagger}\nonumber\\
&&+K_{12} F_0^2\trace{U D^\mu\Udag[\nabla_\mu\qr,\qr] +
             \Udag D^\mu U[\nabla_\mu\ql,\ql]}\\ \nonumber
&&+K_{13} F_0^2\trace{\nabla_\mu\qr\, U \nabla^\mu\ql\,\Udag}\\ \nonumber
&&+K_{14} F_0^2\trace{\nabla_\mu\ql \nabla^\mu\ql +\nabla_\mu\qr \nabla^\mu\qr} 
\ena
where
\be
\nabla_\mu\ql=\partial_\mu\ql-i[l_\mu,\ql],\quad
\nabla_\mu\qr=\partial_\mu\qr-i[l_\mu,\qr] \ .
\en
As usual, the correlators collect the contributions at one-loop 
generated from
${\cal L}^{(2)}_{chir}$ and the tree contributions from 
${\cal L}^{e^2p^2}_{chir}$. Let us first quote the results, some details
about how the calculation proceeds will be given later. For the
components of $<AAQ_AQ_A>$ one obtains
\bea\lbl{kif1}
&&<AAQ_AQ_A>_1=
F_0^2\,[ 2(K^r_1-K^r_2)+2(K^r_5-K^r_6)+4K^r_{12}+K^r_{13}+2K^r_{14}
+\chi_{1}(\mg,\mu) 
\nonumber\\
&&\phantom{ <AAQ_AQ_A>_1=  F_0^2\,[}
+{5\over 4} Z(k,\mu) ]
\nonumber\\
&&<AAQ_AQ_A>_2=F_0^2\,[-2(2K^r_3+K^r_4)+{4\over3}(K^r_5-K^r_6)
+{3\over 2} Z(k,\mu) ]\nonumber\\
&&<AAQ_AQ_A>_3=F_0^2\,[6(K^r_1-K^r_2)+2(K^r_5-K^r_6) 
+{9\over 4} Z(k,\mu) ]
\ena
where $K^r_i$ stand for the renormalized scale dependent parameters.
The function $\chi_1(\mg,\mu)$ is one of $\chi_i$, $i=1,2,3,4$  
which are  the one-loop chiral contributions involving a virtual photon. 
We will describe the computation in detail in the next section. 
The contributions proportional to the function  $Z(k,\mu)$ arise from 
purely pionic one-loop diagrams having one vertex from the term proportional
to $C$ in eq.\rf{lagchir2}.  This function reads
\be
Z(k,\mu)= {2C\over 16\pi^2 F_0^4} \left(-\log{-k^2\over\mu^2}+1\right)\ .
\en
While we include this contribution here for completeness we note that it
is of subleading order in the large $N_c$ expansion. In the following, 
we will  drop such contributions for consistency with the 
leading large $N_c$ approximation which we will make in the treatment of
the resonances.
We also note  that the correlators must be independent of the
chiral renormalization scale $\mu$. This constraint allows one to recover
the scale dependence of $K^r_i(\mu)$ computed 
in refs.\cite{urech}\cite{neufeld-ruper}, 
generalized to an arbitrary $\xi$  gauge.
The components of the second correlator has the following expression 
\bea\lbl{kif2}
&&<AAQ_VQ_V>_1=F_0^2\,[2(K^r_1+K^r_2)+2(K^r_5+K^r_6)
+4K^r_{12}-K^r_{13}+2K^r_{14} +\chi_{2}(\mg,\mu) 
\nonumber\\
&&\phantom{ <AAQ_AQ_A>_1=  F_0^2\,[}
-{5\over 4} Z(k,\mu) ]
\nonumber\\
&&<AAQ_VQ_V>_2=F_0^2\,[-2(2K^r_3-K^r_4)+{4\over3}(K^r_5+K^r_6)
-{3\over 2} Z(k,\mu)]
\nonumber\\
&&<AAQ_VQ_V>_3=F_0^2\,[6(K^r_1+K^r_2)+2(K^r_5+K^r_6)-{9\over 4} Z(k,\mu)]\ .
\ena
For the next two correlators, only the first flavour structure 
appears at this chiral order
\bea\lbl{kif3}
&&<VVQ_AQ_A>_1=F_0^2\,[-K^r_{13}+2K^r_{14}+\chi_{3}(\mg,\mu)]
\nonumber\\
&&<VVQ_VQ_V>_1=F_0^2\,[ K^r_{13}+2K^r_{14}]\ .
\ena
The last correlator has two non-vanishing flavour structures.
The  structure proportional to $\efour$ is the only one to receive a
tree-level contribution from $\lag_{chir}^{(2)}$,
\bea\lbl{kif4}
&&<AVQ_AQ_V>_1=F_0^2\,[2K^r_{12}+2K^r_{14}+\chi_{4}(\mg,\mu)]
\nonumber\\
&&<AVQ_AQ_V>_4={2C\over k^2}+F_0^2\,[-K^r_{13} + 2K^r_{12}+\chi_{4}(\mg,\mu)]\ .
\ena

We now introduce  a set of combinations of these correlator components
which enjoy useful properties
\bea\lbl{combi}
&&\Pi_1=\correl{AAQ_AQ_A}_1-2\correl{AVQ_AQ_V}_1+\correl{VVQ_AQ_A}_1
\nonumber\\
&&\Pi_2=\correl{AAQ_VQ_V}_1-2\correl{AVQ_AQ_V}_1+\correl{VVQ_VQ_V}_1
\nonumber\\
&&\Pi_3=\correl{AAQ_AQ_A}_3
\nonumber\\
&&\Pi_4=\correl{AAQ_VQ_V}_3
\nonumber\\
&&\Pi_5=\correl{AAQ_AQ_A}_2-{2\over3}\correl{AAQ_AQ_A}_3
\nonumber\\
&&\Pi_6=\correl{AAQ_VQ_V}_2-{2\over3}\correl{AAQ_VQ_V}_3 \ .
\ena
These six combinations are in one to one correspondence with the 
parameters $K^r_1$,...,$K^r_6$ which we intend to determine. One can show 
(see \cite{bmdash}) that they are independent of the gauge parameter $\xi$, 
and we will show below that they get no  contributions from the 
QCD counterterms. 
This last fact implies that the photon loop integral must be converging.
Convergence must be understood in the sense that integration 
is to be performed 
in $d$ dimensions and the limit $d\to4$ is well defined. 

In the leading large $N_c$ approximation one has
\be\lbl{bignc}
\Pi_5=\Pi_6=0\quad\quad (\rm{large}\ N_c)\ .
\en
This is not difficult to show. 
The graphs which are of leading order in $N_c$ are the planar ones which
involve a single fermion loop\cite{thooft}. 
This implies that they involve a single
trace over the flavour $\lambda$ matrices. Furthermore,  we are interested
in four-point QCD correlators which are symmetric under permutation of the 
Lorentz indices $\alpha$, $\beta$  as well as in the other two 
indices $\rho\sigma$ (see eq.\rf{convol}). 
They must then have the following structure 
\bea
\correl{A^a_\alpha A^b_\beta Q^c_A Q^d_A}=&&ig_{\alpha\beta}
\left( 
E\trace{\{\lambda^a,\lambda^b\}\{\lambda^c,\lambda^d\}}
+F\trace{\lambda^a\lambda^c\lambda^b\lambda^d + 
         \lambda^a\lambda^d\lambda^b\lambda^c} \right)
\nonumber\\
&&= ig_{\alpha\beta} 
\left( 2(2E-F)\,\eone +{4\over3}(2E+F)(\etwo+{3\over2}\ethree)
\right)
\ena
making use of relations \rf{flavrel}, where $E$ and $F$ are arbitrary,
and the same holds for the propagator with $Q_AQ_A$ replaced by $Q_VQ_V$. 
This proves relations\rf{bignc}. From these, one can recover the large
$N_c$ relations among the corresponding $K^r_i$ parameters 
\be
K^r_3=-K^r_1\quad K^r_4=2K^r_2  \quad\quad (\rm{large}\ N_c)\ 
\en
noted in refs.\cite{urech} and \cite{bij-prad}.
We will further consider the following three combinations  
\bea\lbl{combib}
&&\Pi_7=\correl{VVQ_VQ_V}_1-\correl{VVQ_AQ_A}_1
\nonumber\\
&&\Pi_8=\correl{AVQ_AQ_V}_1-\correl{VVQ_VQ_V}_1
\nonumber\\
&&\Pi_9=\lim_{k^2=0} k^2\correl{AVQ_AQ_V}_4
\ena
From these combinations we will recover established  
results concerning the couplings
$C$, $K^r_{12}$, $K^r_{13}$. $\Pi_7$ and $\Pi_8$ depend on the gauge. 
Convergence
of the photon loop integral can be established for $\Pi_7$ and $\Pi_9$ while
for $\Pi_8$ it holds in the particular gauge $\xi=0$ and in a limited sense, 
as we will discuss in more detail. This limitation affects the
determination of $K^r_{12}$. We will not attempt to estimate the coupling 
$K^r_{14}$ which is a pure source term. 

One further remark is in order. 
One usually expects the coupling constants in ChPT  to be associated with 
QCD correlators which are order parameters, except for the pure source
terms. Looking at eqs.\rf{combi} this, at first sight,  
seems sometimes not to be the case here. 
For instance, $\Pi_3$ is associated with the correlator 
$<A_\alpha^a A_\beta^b A_\mu^c A_\nu^d>$ which does not vanish in 
perturbation theory. However, this
is simply due to the fact that the components
$<AVQ_AQ_V>_3$ and $<VVQ_VQ_V>_3$ vanish at chiral order $e^2p^2$, so one
could equally well express $\Pi_3$ in terms of a correlator which does behave
as an order parameter.

\subsection{Calculation of the one-loop photon contributions} 
In this section, we will compute  contributions to the generating functional
$W(v_\mu,a_\mu,q_V,q_A)$ arising from the chiral Lagrangian 
${\cal L}^{(2)}_{chir}$. The use of the spurions $q_V$, $q_A$ as sources
gives rise to a simple diagrammatic method which we will  
in this section and the following ones.
As there were some errors reported in related calculations in the past
(see e.g.\cite{bu0}) and we will, in fact, find some discrepancies with BU,
we will present our calculations in some detail.
Here, we will compute the functions $\chi_i(\mg,\mu)$ 
which appear in eqs.\rf{kif2}. They arise from sets of one-loop diagrams
containing one photon line and one pion line, the vertices being 
obtained from the leading order Lagrangian \rf{lagchir2}. 
The corresponding 
standard $\xi$-gauge massive photon propagator reads
\be\lbl{photprop}
D_F^{\mu\nu}(p)={-i\over (p^2-\mgd)} \left\{ g^{\mu\nu} +(\xi-1)
{p^\mu p^\nu\over (p^2-\xi\mgd)}\right\}\ .
\en
The list of the vertices which are effectively needed is displayed in
the appendix.
We  remark here that these contributions are to be included both 
in the chiral expansion of the correlator as well as in the expression
from the resonance Lagrangian since the latter includes the piece from
eq.\rf{lagchir2}. 

Let us now compute the contribution to $\correl{AAQ_AQ_A}_i$. A single diagram
contributes, shown  below
\begin{center}
\begin{texdraw}
\drawdim cm 
\move(0 0)
\textref h:C v:C \htext(1.0 3.0){\qa}
\textref h:C v:C \htext(0.9 1.5){\amu}
\textref h:C v:C \htext(1.2 1.5){\amu}
\textref h:C v:C \htext(1.0 0.0){\qa}
\move(1 0) \linewd 0.04 \setgray 0 \lvec(1 3)
\end{texdraw}
\end{center}
The following conventions are used to represent the sources
\begin{center}
\begin{texdraw}
\drawdim cm 
\setgray 1
\textref h:C v:C \htext(1.0 1.0){\qa=}
\textref h:L v:C \htext(1.8 1.0){$q_A$ source}
\textref h:C v:C \htext(1.0 0.0){\qv=}
\textref h:L v:C \htext(1.8 0.0){$q_V$ source}
\textref h:C v:C \htext(7.0 1.0){\amu=}
\textref h:L v:C \htext(7.8 1.0){$a_\mu$ source}
\textref h:C v:C \htext(7.0 0.0){\vmu=}
\textref h:L v:C \htext(7.8 0.0){$v_\mu$ source}
\end{texdraw}
\end{center}
The relevant vertices are listed in the appendix.
One must keep in mind, for instance,
that the coupling of one axial source to the pion does not contribute
to our correlators. 
The photon line which joins the two $q$-sources is 
not explicitly drawn. 
Each diagram is easy to interpret in terms of the integral 
representation eq.\rf{convol}.
We will first give the results in integral form, in
order to display the integrands, in which we perform the replacement of
$p_\alpha p_\beta$  by $p^2 g_{\alpha\beta}/ d$.
In integral form, this diagrams gives
\be
\correl{AAQ_AQ_A}_1=-F_0^2 \int \iddp {\xi\over p^2 (p^2-\xi \mgd)}
\en
and the other components $\correl{AAQ_AQ_A}_i$ get no contributions.
Next, we consider $\correl{VVQ_VQ_V}_1$  and, again, there is a single 
diagram to compute shown below
\begin{center}
\begin{texdraw}
\drawdim cm 
\move(1 0)
\textref h:C v:C \htext(1.2 3.0){\qv}
\textref h:C v:C \htext(0.9 3.0){\amu}
\textref h:C v:C \htext(1.2 0.0){\qv}
\textref h:C v:C \htext(0.9 0.0){\amu}
\move(1 0) \linewd 0.04 \setgray 0 \lvec(1 3)
\end{texdraw}
\end{center}
which gives
\be
\correl{AAQ_VQ_V}_1=-F_0^2 \int \iddp \left\{
{1\over p^2(p^2-\mgd)}+{1\over d}{\xi-1\over (p^2-\mgd)(p^2-\xi\mgd)}\right\}
\en
and the other components vanish. One also finds that $<VVQ_VQ_V>_i$ receive
no contribution from pions. Next we consider $<VVQ_AQ_A>_i$, in this
case there are four diagrams
\begin{center}
\begin{texdraw}
\drawdim cm 
\move(1 0)
\textref h:C v:C \htext(1.2 3.0){\qa}
\textref h:C v:C \htext(0.9 3.0){\vmu}
\textref h:C v:C \htext(0.9 0.0){\vmu}
\textref h:C v:C \htext(1.2 0.0){\qa}
\move(1 0) \linewd 0.04 \setgray 0 \lvec(1 3)
\textref h:C v:C \htext(4.0 3.0){\qa}
\textref h:C v:C \htext(3.9 1.5){\vmu}
\textref h:C v:C \htext(4.2 1.5){\vmu}
\textref h:C v:C \htext(4.0 0.0){\qa}
\move(4 0) \linewd 0.04 \setgray 0 \lvec(4 3)
\textref h:C v:C \htext(7.0 3.0){\qa}
\textref h:C v:C \htext(7.0 1.0){\vmu}
\textref h:C v:C \htext(7.0 2.0){\vmu}
\textref h:C v:C \htext(7.0 0.0){\qa}
\move(7 0) \linewd 0.04 \setgray 0 \lvec(7 3)
\textref h:C v:C \htext(10.0 3.0){\qa}
\textref h:C v:C \htext(10 1.5){\vmu}
\textref h:C v:C \htext(10.2 0.0){\qa}
\textref h:C v:C \htext(10.0 0.0){\vmu}
\move(10 0) \linewd 0.04 \setgray 0 \lvec(10 3)
\end{texdraw}
\end{center}
Calculating these diagrams, one finds that the last two cancel each other and 
the first two give  the following integral
\be
<VVQ_AQ_A>_1= F_0^2 \left(1-{1\over d}\right) \int \iddp { \xi-1\over 
(p^2-\mgd)(p^2-\xi\mgd) }\ .
\en
Finally, we have to compute $<AVQ_AQ_V>_i$. There are four diagrams
in this case

\begin{center}
\begin{texdraw}
\drawdim cm 
\textref h:C v:C \htext(1.0 3.0){\qa}
\textref h:C v:C \htext(1.0 1.5){\vmu}
\textref h:C v:C \htext(1.2 0.0){\qv}
\textref h:C v:C \htext(1.0 0.0){\amu}
\move(1 0) \linewd 0.04 \setgray 0 \lvec(1 3)
\textref h:C v:C \htext(4.2 3.0){\qa}
\textref h:C v:C \htext(4.0 3.0){\vmu}
\textref h:C v:C \htext(4.2 0.0){\qv}
\textref h:C v:C \htext(4.0 0.0){\amu}
\move(4 0) \linewd 0.04 \setgray 0 \lvec(4 3)
\textref h:C v:C \htext(7.2 3.0){\amu}
\textref h:C v:C \htext(7.0 3.0){\vmu}
\textref h:C v:C \htext(7.0 1.5){\qv}
\textref h:C v:C \htext(7.0 0.0){\qa}
\move(7 0) \linewd 0.04 \setgray 0 \lvec(7 3)
\textref h:C v:C \htext(10.2 3.0){\amu}
\textref h:C v:C \htext(10.0 3.0){\vmu}
\textref h:C v:C \htext(10.2 0.0){\qv}
\textref h:C v:C \htext(10.0 0.0){\qa}
\move(10 0) \linewd 0.04 \setgray 0 \lvec(10 3)\end{texdraw}
\end{center}

Computing these diagrams gives
\bea
&&<AVQ_AQ_V>_1=-F_0^2\int \iddp\Bigg\{
{1\over d}{\xi\over p^2(p^2-\xi \mgd)}
-{1\over 2}{1\over p^2(p^2- \mgd)}
\nonumber\\
&&\phantom{<AVQ_AQ_V>_1=-F_0^2 }
-{1\over 2d}{\xi-1\over (p^2- \mgd)(p^2-\xi \mgd)} \Bigg\}
\nonumber\\
&& <AVQ_AQ_V>_4={F_0^2(1-d)\over k^2}\int\iddp {1\over p^2}+<AVQ_AQ_V>_1
\ena
The $1/k^2$ term is generated by the third and fourth diagrams and the integral
vanishes in dimensional regularization.
We can now compute the d-dimensional integrals and the divergences are removed
using the chiral $\overline{MS}$ prescription\cite{gl85}.
The results for the loop functions $\chi_i(\mg,\mu)$ are collected below
\bea\lbl{loopres}
&&\chi_1(\mg,\mu)= {1\over16\pi^2}
\left[\xi\lngmu +\xi\log\xi \right]
\nonumber\\
&&\chi_2(\mg,\mu)={1\over16\pi^2}\left({1\over8}\right)
\left[2(\xi+3)\lngmu +2\xi\log\xi -\xi+1\right]
\nonumber\\
&&\chi_3(\mg,\mu)={1\over16\pi^2}\left(-{1\over8}\right)
\left[6(\xi-1)\lngmu +6\xi\log\xi +\xi-1\right]
\nonumber\\
&&\chi_4(\mg,\mu)={1\over16\pi^2}\left({1\over16}\right)
\left[2(\xi-3)\lngmu +2\xi\log\xi -\xi-1\right]\ .
\ena
Using these results and those from the preceding section which display
the contributions involving the chiral couplings $K^r_i$ we can deduce the
scale and gauge dependence of the latter. One finds that $K^r_1$,...,$K^r_6$
are independent of the gauge parameter $\xi$ and the scale dependence is in
agreement with the results of \cite{urech,neufeld-ruper}. 
The couplings $K^r_{12}$, $K^r_{13}$, $K^r_{14}$
do depend on $\xi$.  Defining the beta functions $\Sigma_i$ in the conventional
manner
\be
{dK^r_i(\mu)\over d\mu}\equiv -{1\over16\pi^2}\,\Sigma_i
\en
one finds from  eqs.\rf{loopres}
\be
\Sigma_{12}= {1\over4}(3-2\xi),\quad
\Sigma_{13}= {3\over4}(1- \xi),\quad
\Sigma_{14}=-{3\over8}(1- \xi)\ .
\en
These results agree with those of 
Urech in the special case $\xi=1$, and they reproduce
the results presented in ref.\cite{bmdash} where a different set of 
correlation functions were employed.

\section{Correlators from the minimal resonance Lagrangian}\label{sec:minreslag}
\subsection{$O(p^4)$ resonance Lagrangian}
Our starting point for determining  the $K^r_i$ parameters is the
integral representation eq.\rf{convol} in terms of 4-point
correlation functions. The first approximation that we make is to consider
the leading large $N_c$ limit. In this limit, it is
an exact statement that  QCD correlators
can be expressed as a sum over tree graphs involving 
resonance propagators\cite{thooft} (see e.g.\cite{derafael} for a review of
recent applications). 
The second approximation that we make is to restrict
ourselves to contributions from a finite number of resonances. This is
acceptable in our case because we consider combinations
of correlators which fall off as powers for large momenta. 

In order to guarantee the validity of chiral Ward identities one may
compute the correlators starting from a chiral Lagrangian with resonances.
In cases where the integrals can be shown to converge, 
imposing the convergence property 
generates a set of constraints for the parameters of the resonance Lagrangian
which we will call consistency conditions. 
These conditions will be discussed in detail in the following.
We begin  with the resonance
Lagrangian which was used by Baur and Urech\cite{bu}
\be\lbl{lagres}
{\cal L}^{(4)}_{res}={\cal L}^{p^2}_{chir}+{\cal L}^{(4)}_{S,{S_0}}
+{\cal L}^{(4)}_{V,A}\ .
\en
BU have used exactly the Lagrangian proposed by Ecker et al.\cite{egpr}
(except for leaving  out the $\pi'$ resonance multiplet)
which we will call the minimal resonance Lagrangian. 
Let us recall the notation and formalism used in ref.\cite{egpr}. 
The  formalism that is used is one
in which all resonances transform homogeneously under the non linear 
representation of the chiral group including the vector and the axial-vector 
resonances which are described as antisymmetric tensor fields. 
In this approach, all the resonance fields can be assigned a chiral
order equal to two, such that the resonance Lagrangian of ref.\cite{egpr},
contains all possible  terms, for each resonance field, 
which are of chiral order four. The only terms 
of chiral order six which are present are the kinetic energy terms. 

We recall here the details: in the scalar sector an octet $S$ and a 
singlet $S_0$ are considered and 
the Lagrangian  involves two coupling-constants,
$c_d$ and $\tilde c_d$,
\bea\lbl{lags4}
&&{\cal L}_{S,S_0}={\undemi}\trace{\nabla^\lambda S \nabla_\lambda S-M^2_S S^2}
+\undemi[\partial^\lambda S_0 \partial_\lambda S_0 - M^2_{S_0}S^2_0 ]
\nonumber\\
&&+c_d\trace{ S u_\mu u^\mu} + \tilde c_d S_0\trace{u_\mu u^\mu}\ .
\ena
Next, a nonet of  vector and axial-vector resonances are considered 
and the minimal Lagrangian
involves three coupling constants $F_A$, $F_V$ and $G_V$,
\bea\lbl{lagav4}
&&{\cal L}_{V,A}=-\undemi\sum_{R=V,A}\trace{ \nabla^\lambda R_{\lambda\mu}
\nabla_\nu R^{\nu\mu}-\undemi M^2_R R_{\mu\nu}R^{\mu\nu} }\nonumber\\
&&{F_A\over2\sqrt2}\trace{A_{\mu\nu}f^{\mu\nu}_-  }+
  {F_V\over2\sqrt2}\trace{V_{\mu\nu}f^{\mu\nu}_+  }+
 {iG_V\over2\sqrt2}\trace{V_{\mu\nu}[u^\mu,u^\nu] }\ .
\ena
We will  compute the basic correlators introduced in sec.1 
first from this minimal resonance Lagrangian.  
It is convenient to keep using the $q_V$, $q_A$ source spurions at this level,
introduced in association with the $v_\mu$ and $a_\mu$ sources, 
with  the replacements
$l_\mu\to l_\mu +q_L F_\mu$, $r_\mu\to r_\mu +q_R F_\mu$.
The diagrammatic calculation is exactly equivalent to using the 
convolution representation
eq.\rf{convol} in which the QCD correlator is computed from the resonance
Lagrangian. In principle, we should be able to reproduce the results
of BU (who used a completely different  approach to
determine the $K^r_i$'s ). We follow their convention to regularize the UV
divergences in the chiral $\overline{MS}$ prescription and we introduce
a regularization scale $\Lambda$. 
The various vertices needed in the calculation are collected in the appendix.
We will use the photon propagator in a general $\xi$
gauge as  in eq.\rf{photprop}. Since the resonances are massive one can
set the photon mass $\mg=0$ in the following.
Since the $O(p^2)$ chiral Lagrangian is also part of the resonance Lagrangian,
the pion contributions
$\chi_i(\mg,\Lambda)$, $i=1,2,3,4$  which we already computed 
(eqs.\rf{loopres}) must be added to the resonance contributions.

\subsection{Contributions proportional to $F_A^2$ and $F_V^2$}
We will now consider successively the contributions proportional to $F_A^2$,
$F_V^2$, $G_VF_V$, $G^2_V$, $G^2_VF_V^2$ which are the only ones 
generated from the vector and axial-vector part of the resonance Lagrangian
eq.\rf{lagres}. 
One easily sees, at first,  
that the following correlator components $\correl{AAQ_AQ_A}_i $
and $\correl{VVQ_VQ_V}_1$ get no contributions
proportional to $F_A^2$.
Next, one finds that there are five diagrams which contribute to 
$\correl{VVQ_AQ_A}_i$, the relevant vertices can be found in appendix A. 
The diagram shown below is found to make
a vanishing contribution
\begin{center}
\begin{texdraw}
\drawdim cm 
\move(0 0)
\textref h:C v:B \htext(1.0 3.0){\qa}
\textref h:C v:C \htext(0.9 1.5){\vmu}
\textref h:C v:C \htext(1.2 1.5){\vmu}
\textref h:C v:C \htext(1.0 0.0){\qa}
\textref h:R v:C \htext(0.6 3.0){$F_A$}
\textref h:R v:B \htext(0.6 0.0){$F_A$}
\move(1 0) \linewd 0.20 \setgray 0.8 \lvec(1 1.5)
           \linewd 0.04 \setgray 0.0 \lvec(1 3)
\end{texdraw}
\end{center}
as it is antisymmetric in the $\alpha$ $\beta$ indices. 
The remaining four diagrams are
\begin{center}
\begin{texdraw}
\drawdim cm 
\move(0 0)
\textref h:C v:C \htext(1.0 3.0){\qa}
\textref h:C v:C \htext(1.0 2.0){\vmu}
\textref h:C v:C \htext(1.0 1.0){\vmu}
\textref h:C v:C \htext(1.0 0.0){\qa}
\textref h:R v:C \htext(0.6 3.0){$F_A$}
\textref h:R v:B \htext(0.6 0.0){$F_A$}
\move(1 0)\linewd 0.20 \setgray 0.8 \lvec(1 3)
\textref h:C v:C \htext(4.0 3.0){\qa}
\textref h:C v:C \htext(3.9 1.5){\vmu}
\textref h:C v:C \htext(4.2 1.5){\vmu}
\textref h:C v:C \htext(4.0 0.0){\qa}
\move(4 0) \linewd 0.20 \setgray 0.8 \lvec(4 3)
\textref h:C v:C \htext(7.2 3.0){\qa}
\textref h:C v:C \htext(6.9 3.0){\vmu}
\textref h:C v:C \htext(6.9 0.0){\vmu}
\textref h:C v:C \htext(7.2 0.0){\qa}
\move(7 0) \linewd 0.20 \setgray 0.8 \lvec(7 3)

\textref h:C v:C \htext(10 3.0){\qa}
\textref h:C v:C \htext(10 1.5){\vmu}
\textref h:C v:C \htext(9.9 0.0){\vmu}
\textref h:C v:C \htext(10.2 0.0){\qa}
\move(10 0)\linewd 0.20 \setgray 0.8 \lvec(10 3)
\end{texdraw}
\end{center}
The result is of the following form
\be
\correl{VVQ_AQ_A}_1=F_A^2\int {\iddp}\sum_1^4 D_i(p)\ ,
\en
with the individual diagrams contributions to the integrand reading
\bea
&&D_1=(1-{1\over d})\left(
 {2-d\over \mad(p^2-\mad)}
+{6-d\over     (p^2-\mad)^2 }
+{4\mad\over   (p^2-\mad)^3 }\right)
\nonumber\\
&&D_2=(1-{1\over d}){-2 \over     (p^2-\mad)^2 }
\nonumber\\
&&D_3=(1-{1\over d})\left({2-d\over\mad(p^2-\mad)}
+ {d-1+\xi\over p^2(p^2-\mad)}  \right)
\nonumber\\
&&D_4=(1-{1\over d})\left(
 {2(d-2)\over \mad(p^2-\mad)} 
-{4     \over     (p^2-\mad)^2 }\right) \ ,
\ena
respectively. Next, we consider $<AA Q_VQ_V>$. A single diagram contributes 
(with a crossed one) which gives the same result as $D_3$ above.
Finally, we turn to $<AV Q_AQ_V>$ . There are three diagrams which
contribute (and no crossed diagrams in this case), which are shown below.
\begin{center}
\begin{texdraw}
\drawdim cm 
\move(0 0)
\move(1 0)
\textref h:C v:C \htext(1.2 3.0){\qa}
\textref h:C v:C \htext(0.9 3.0){\vmu}
\textref h:C v:C \htext(0.9 0.0){\amu}
\textref h:C v:C \htext(1.2 0.0){\qv}
\textref h:R v:C \htext(0.6 3.0){$F_A$}
\textref h:R v:B \htext(0.6 0.0){$F_A$}
\linewd 0.20
\setgray 0.8
\move(1 0)  \lvec(1 3)
\move(4 0)
\textref h:C v:C \htext(4 3.0){\qa}
\textref h:C v:C \htext(4 1.5){\vmu}
\textref h:C v:C \htext(3.9 0.0){\amu}
\textref h:C v:C \htext(4.2 0.0){\qv}
\linewd 0.20
\setgray 0.8
\move(4 0)  \lvec(4 3)
\move(7 0)
\textref h:C v:B \htext(6.9 3.0){\amu}
\textref h:C v:C \htext(7.2 3.0){\vmu}
\textref h:C v:C \htext(7.0 1.5){\qv}
\textref h:C v:C \htext(7.0 0.0){\qa}
\textref h:R v:C \htext(6.6 1.5){$F_A$}
\textref h:R v:B \htext(6.6 0.0){$F_A$}
\linewd 0.20
\setgray 0.8
\move(7 0)  \lvec(7 1.5)
\linewd 0.04  \setgray 0.0  \lvec(7 3)
\end{texdraw}
\end{center}
These diagrams give
\bea
&&\correl{AVQ_AQ_V}_1=F_A^2(1-{1\over d})\int\iddp\left[ {d+\xi-1\over 2p^2
(p^2-\mad)}-{1\over(p^2-\mad)^2}\right]
\nonumber\\
&& k^2\correl{AVQ_AQ_V}_4=F_A^2\int \iddp{1-d\over p^2-\mad}\ .
\ena
Computing all the integrals, in summary, the following results are
obtained 
\bea
&&\correl{AAQ_AQ_A}^\fad_i=0
\nonumber\\
&&\correl{AAQ_VQ_V}^\fad_1=
-{3\over 4}{F_A^2\over 16\pi^2}
(1+\xi)
\left({1\over6}+\log {M^2_A\over\Lambda^2} \right)
\nonumber\\
&&\correl{VVQ_AQ_A}^\fad_1=
{3\over 2}{F_A^2\over 16\pi^2}
\left\{ 1-{1\over2}(\xi-1)\left({1\over6}+\log {M^2_A\over\Lambda^2} 
\right)\right\}
\nonumber\\
&&\correl{VVQ_VQ_V}^\fad_1=0
\nonumber\\
&&\correl{A VQ_AQ_V}^\fad_1= 
-{3\over8}{F_A^2\over 16\pi^2} 
(1+\xi)
\left({1\over6}+\log {M^2_A\over\Lambda^2} \right)
\nonumber\\
&&k^2 \correl{A VQ_AQ_V}^\fad_4= 
3{M^2_A F_A^2 \over 16\pi^2}  
( {2\over3} + \log{M^2_A\over\Lambda^2} ) + O(k^2)\ .
\ena
There is a simple correspondence between 
the diagrams proportional to $F_V^2$
and those proportional to $F_A^2$ considered above so we give directly
the results for the various components of the correlators
\bea
&&\correl{AAQ_AQ_A}^\fvd_1=-{3\over 4}{F_V^2\over 16\pi^2}
(1+\xi)
\left({1\over6}+\log {M^2_V\over\Lambda^2} \right)
\nonumber\\
&&\correl{AAQ_VQ_V}^\fvd_i=0\nonumber\\
&&\correl{VVQ_AQ_A}^\fvd_1=0\nonumber\\
&&\correl{VVQ_VQ_V}^\fvd_1=
{3\over 2} {F_V^2\over 16\pi^2}
\left\{ 1-{1\over2}(\xi-1)\left({1\over6}+\log {M^2_V\over\Lambda^2} \right)
\right\}
\nonumber\\
&&\correl{AVQ_AQ_V}^\fvd_1=
-{3\over8}{F_V^2\over 16\pi^2} 
(1+\xi)
\left({1\over6}+\log {M^2_V\over\Lambda^2} \right)
\nonumber\\
&&k^2 \correl{AVQ_AQ_V}^\fvd_4=
- 3{M^2_V F_V^2 \over 16\pi^2}  
( {2\over3} + \log{M^2_V\over\Lambda^2} ) \ .
\ena

\subsection{$G_VF_V$ contributions }
We have two diagrams which contribute to $\correl{AAQ_AQ_A}_1$ shown below.

\begin{center}
\begin{texdraw}
\drawdim cm 
\move(0 0)
\textref h:C v:C \htext(1.2 3.0){\qa}
\textref h:C v:C \htext(0.9 3.0){\amu}
\textref h:C v:C \htext(0.9 0.0){\amu}
\textref h:C v:C \htext(1.2 0.0){\qa}
\textref h:R v:C \htext(0.6 3.0){$G_V$}
\textref h:R v:B \htext(0.6 0.0){$F_V$}
\move(1 0) \linewd 0.20 \setgray 0.5 \lvec(1 3)
\textref h:C v:C \htext(4.0 3.0){\qa}
\textref h:C v:C \htext(3.9 2.0){\amu}
\textref h:C v:C \htext(3.9 0.0){\amu}
\textref h:C v:C \htext(4.2 0.0){\qa}
\textref h:R v:C \htext(3.6 2.0){$G_V$}
\textref h:R v:B \htext(3.6 0.0){$F_V$}
\move(4 0) \linewd 0.20 \setgray 0.5 \lvec(4 2)
           \linewd 0.04 \setgray 0.0 \lvec(4 3)
\end{texdraw}
\end{center}
This results in the following integral for the correlator
\be
\correl{AAQ_AQ_A}_1=4\, G_VF_V(1-{1\over d})\int
\iddp \left( {d-2\over\mvd(p^2-\mvd)}+{1-d\over p^2(p^2-\mvd)}\right)\ .
\en
Next we have one diagram contributing to $\correl{AAQ_VQ_V}_1$ 
which is non vanishing
\begin{center}
\begin{texdraw}
\drawdim cm 
\move(0 0)
\textref h:C v:C \htext(1.2 3.0){\qv}
\textref h:C v:C \htext(0.9 3.0){\amu}
\textref h:C v:C \htext(0.9 2.0){\amu}
\textref h:C v:C \htext(1.2 0.0){\qv}
\textref h:R v:C \htext(0.6 2.0){$G_V$}
\textref h:R v:B \htext(0.6 0.0){$F_V$}
\move(1 0)\linewd 0.20 \setgray 0.5 \lvec(1 2)
          \linewd 0.04 \setgray 0.0 \lvec(1 3)
\end{texdraw}
\end{center}
and which gives
\be
\correl{AAQ_VQ_V}_1=4G_VF_V\int\iddp {1\over p^2(p^2-\mvd)}\ .
\en
The two correlators $\correl{VVQ_AQ_A}_1$ and 
$\correl{VVQ_VQ_V}_1$ are then found to get no contributions proportional to
$G_VF_V$.

Next, we consider $\correl{AVQ_AQ_V}_i$. One first finds a set of 7
diagrams which contribute both to $\correl{AVQ_AQ_V}_1$ and 
$\correl{AVQ_AQ_V}_4$. The three diagrams drawn below give zero
contribution
\begin{center}
\begin{texdraw}
\drawdim cm 
\move(0 0)
\textref h:C v:C \htext(1.0 3.0){\qa}
\textref h:C v:C \htext(0.9 2.0){\amu}
\textref h:C v:C \htext(1.2 2.0){\vmu}
\textref h:C v:C \htext(1.0 0.0){\qv}
\textref h:R v:C \htext(0.6 2.0){$G_V$}
\textref h:R v:B \htext(0.6 0.0){$F_V$}
\move(1 0)\linewd 0.20 \setgray 0.5 \lvec(1 2)
          \linewd 0.04 \setgray 0.0 \lvec(1 3)
\textref h:C v:C \htext(4.0 3.0){\qa}
\textref h:C v:C \htext(3.9 2.0){\amu}
\textref h:C v:C \htext(4.2 2.5){\vmu}
\textref h:C v:C \htext(4.0 0.0){\qv}
\move(4 0)\linewd 0.20 \setgray 0.5 \lvec(4 2)
          \linewd 0.04 \setgray 0.0 \lvec(4 3)
\textref h:C v:C \htext(7.0 3.0){\qa}
\textref h:C v:C \htext(6.9 2.0){\amu}
\textref h:C v:C \htext(7.0 1.0){\vmu}
\textref h:C v:C \htext(7.0 0.0){\qv}
\move(7 0)\linewd 0.20 \setgray 0.5 \lvec(7 2)
          \linewd 0.04 \setgray 0.0 \lvec(7 3)
\end{texdraw}
\end{center}
The next four diagrams are
\begin{center}
\begin{texdraw}
\drawdim cm 
\move(0 0)
\textref h:C v:C \htext(1.2 3.0){\qa}
\textref h:C v:C \htext(0.9 3.0){\amu}
\textref h:C v:C \htext(0.9 0.0){\vmu}
\textref h:C v:C \htext(1.2 0.0){\qv}
\textref h:R v:C \htext(0.6 3.0){$G_V$}
\textref h:R v:B \htext(0.6 0.0){$F_V$}
\move(1 0)\linewd 0.20 \setgray 0.5 \lvec(1 3)
\textref h:C v:C \htext(4.2 3.0){\qa}
\textref h:C v:C \htext(3.9 3.0){\amu}
\textref h:C v:C \htext(3.9 2.0){\vmu}
\textref h:C v:C \htext(4.2 0.0){\qv}
\move(4 0)\linewd 0.20 \setgray 0.5 \lvec(4 3)
\textref h:C v:C \htext(7.0 3.0){\qa}
\textref h:C v:C \htext(6.9 2.0){\amu}
\textref h:C v:C \htext(6.9 0.0){\vmu}
\textref h:C v:C \htext(7.2 0.0){\qv}
\textref h:R v:C \htext(6.6 2.0){$G_V$}
\textref h:R v:B \htext(6.6 0.0){$F_V$}
\move(7 0)\linewd 0.20 \setgray 0.5 \lvec(7 2)
          \linewd 0.04 \setgray 0.0 \lvec(7 3)
\textref h:C v:C \htext(10.2 3.0){\qa}
\textref h:C v:C \htext(10.0 3.0){\vmu}
\textref h:C v:C \htext(9.9  2.0){\amu}
\textref h:C v:C \htext(10.2 0.0){\qv}
\move(10 0)\linewd 0.20 \setgray 0.5 \lvec(10 2)
           \linewd 0.04 \setgray 0.0 \lvec(10 3)
\end{texdraw}
\end{center}
Computing these four diagrams yields the following result for 
$\correl{AVQ_AQ_V}_1$
\be
\correl{AVQ_AQ_V}_1=G_VF_V(1-{1\over d})\int \iddp\left(
{-d\over p^2(p^2-\mvd)}+{2\over (p^2-\mvd)^2}\right)\ ,
\en
these diagrams make an identical contribution also to $\correl{AVQ_AQ_V}_4$.
In order to get the full contribution to $\correl{AVQ_AQ_V}_4$ we must
also add the following two diagrams
\begin{center}
\begin{texdraw}
\drawdim cm 
\move(0 0)
\textref h:C v:C \htext(1.2 3.0){\vmu}
\textref h:C v:C \htext(1.0 3.0){\amu}
\textref h:C v:C \htext(0.9 2.0){\qa}
\textref h:C v:C \htext(1.0 0.0){\qv}
\textref h:R v:C \htext(0.6 2.0){$G_V$}
\textref h:R v:B \htext(0.6 0.0){$F_V$}
\move(1 0)\linewd 0.20 \setgray 0.5 \lvec(1 2)
          \linewd 0.04 \setgray 0.0 \lvec(1 3)
\textref h:C v:C \htext(5.2 3.0){\vmu}
\textref h:C v:C \htext(5.0 3.0){\amu}
\textref h:C v:C \htext(4.0 2.0){\qa}
\textref h:C v:C \htext(5.0 0.0){\qv}
\move(5 0)\linewd 0.20 \setgray 0.5 \lvec(5 2)
          \linewd 0.04 \setgray 0.0 \lvec(4 2)
\move(5 2)\linewd 0.04 \setgray 0.0 \lvec(5 3)
\end{texdraw}
\end{center}
These diagrams are slightly more complicated to compute because one must
allow for a small momentum to flow through the $v_\beta$ source and expand
the result in powers of $k^2$. One finds that the $k^2$ pole vanishes 
and then 
adding the constant contribution to that arising from the preceding four
diagrams one finds that they cancel. In summary, the contributions
proportional to $G_VF_V$ which are non vanishing read
\bea
&&\correl{AAQ_AQ_A}^\gvfv_1={3\,G_VF_V\over 16\pi^2} 
\left({1\over6}+\log {M^2_V\over\Lambda^2} \right)
\nonumber\\
&&\correl{AAQ_VQ_V}^\gvfv_1=-{3\,G_VF_V\over 16\pi^2} 
\left({1\over6}+\log {M^2_V\over\Lambda^2} \right)
\nonumber\\
&&\correl{AVQ_AQ_V}^\gvfv_1={3\over2}\,{G_VF_V\over 16\pi^2} 
\left({1\over6}+\log {M^2_V\over\Lambda^2} \right)\ .
\ena

\subsection{$G^2_V$ and $G^2_V F_V^2$ contributions}
These contributions are simple because a single correlator component
is found to be non vanishing: $\correl{AAQ_AQ_A}_1$ in the case of  
$G^2_V$ and $\correl{AAQ_VQ_V}_1$ in the case of $G^2_V F_V^2$.
There are three diagrams proportional to $G^2_V$ 
\begin{center}
\begin{texdraw}
\drawdim cm 
\move(0 0)
\textref h:C v:C \htext(0.9 3.0){\qa}
\textref h:C v:C \htext(1.1 3.0){\amu}
\textref h:C v:C \htext(0.9 0.0){\amu}
\textref h:C v:C \htext(1.1 0.0){\qa}
\textref h:R v:C \htext(0.6 3.0){$G_V$}
\textref h:R v:B \htext(0.6 0.0){$G_V$}
\linewd 0.20
\setgray 0.5
\move(1.0 0.0)  \lvec(1 3)
\move(4 0)
\textref h:C v:C \htext(4.0 3.0){\qa}
\textref h:C v:C \htext(4.0 2.0){\amu}
\textref h:C v:C \htext(3.9 0.0){\amu}
\textref h:C v:C \htext(4.1 0.0){\qa}
\textref h:R v:C \htext(3.6 2.0){$G_V$}
\textref h:R v:B \htext(3.6 0.0){$G_V$}
\linewd 0.20
\setgray 0.5
\move(4.0 0.0)  \lvec(4 2)
\linewd 0.04  \setgray 0  \lvec(4 3)
\move(7 0)
\textref h:C v:C \htext(7.0 3.0){\qa}
\textref h:C v:C \htext(7.0 2.0){\amu}
\textref h:C v:C \htext(7.0 1.0){\amu}
\textref h:C v:C \htext(7.0 0.0){\qa}
\textref h:R v:C \htext(6.6 2.0){$G_V$}
\textref h:R v:B \htext(6.6 1.0){$G_V$}
\move(7 0) \lvec (7 1)
\move(7 2) \lvec (7 3)
\linewd 0.2 \setgray 0.5
\move(7 1) \lvec(7 2)
\end{texdraw}
\end{center}
and a single one proportional to $G^2_V F_V^2$

\begin{center}
\begin{texdraw}
\drawdim cm 
\move(0 0)
\textref h:C v:C \htext(1.0 3.0){\qv}
\textref h:C v:C \htext(1.0 2.0){\amu}
\textref h:C v:C \htext(1.0 1.0){\amu}
\textref h:C v:C \htext(1.0 0.0){\qv}
\textref h:R v:C \htext(0.6 2.0){$G_V$}
\textref h:R v:B \htext(0.6 1.0){$G_V$}
\textref h:R v:B \htext(0.6 3.0){$F_V$}
\textref h:R v:B \htext(0.6 0.0){$F_V$}
\linewd 0.20 \setgray 0.5
\move(1 0.1) \lvec (1 1)
\move(1 2) \lvec (1 2.9)
\linewd 0.04 \setgray 0
\move(1 1) \lvec(1 2)
\end{texdraw}
\end{center}
After a small computation the following results are obtained,
\bea
&&<AAQ_AQ_A>_1 =4G^2_V (1-{1\over d})\int{\iddp}
\left({2-d\over\mvd (p^2-\mvd)}
+{d-1\over p^2(p^2-\mvd)} \right)
\nonumber\\
&&<AAQ_VQ_V>_1 =-4{G^2_V F_V^2\over F_0^2}(1-{1\over d})\int{\iddp}
{1\over (p^2-\mvd)^2 }\ .
\ena
In summary, the non vanishing contributions proportional to $G_V^2$ read
\bea
&&\correl{AAQ_AQ_A}^{\gvd}_1 =
-{3G^2_V\over 16\pi^2}\left({1\over6} +\log{\mvd\over\Lambda^2}\right)
\nonumber\\
&&\correl{AAQ_VQ_V}^{\gvd\fvd}_1 =
{ 3G^2_VF_V^2\over 16\pi^2 F_0^2} \left({7\over6} 
+\log{\mvd\over\Lambda^2}\right)\ .
\ena

\subsection{Contributions proportional to $c_d$, $\tilde c_d$}
The scalar singlet turns out to contribute only to $\correl{AAQ_AQ_A}_2$ and
the scalar octet to $\correl{AAQ_AQ_A}_3$. The three Feynman diagrams to be
computed are depicted below. 
\begin{center}
\begin{texdraw}
\drawdim cm 
\move(0 0)
\textref h:C v:C \htext(1.2 3.0){\qa}
\textref h:C v:C \htext(0.9 3.0){\amu}
\textref h:C v:C \htext(0.9 0.0){\amu}
\textref h:C v:C \htext(1.2 0.0){\qa}
\move(1 0) \linewd 0.20 \setgray 0.2 \lvec(1 3)
\textref h:C v:C \htext(4.0 3.0){\qa}
\textref h:C v:C \htext(3.9 2.0){\amu}
\textref h:C v:C \htext(3.9 0.0){\amu}
\textref h:C v:C \htext(1.2 0.0){\qa}
\move(4 0) \linewd 0.20 \setgray 0.2 \lvec(4 2)
           \linewd 0.04 \setgray 0.0 \lvec(4 3)
\move(7 0)
\textref h:C v:C \htext(7.0 3.0){\qa}
\textref h:C v:C \htext(6.9 2.0){\amu}
\textref h:C v:C \htext(6.9 1.0){\amu}
\textref h:C v:C \htext(7.0 0.0){\qa}
\move(7 0) 
\linewd 0.04 \setgray 0.0 \lvec(7 1)
\linewd 0.20 \setgray 0.2 \lvec(7 2)
\linewd 0.04 \setgray 0.0 \lvec(7 3)
\end{texdraw}
\end{center}
The dependence on the photon gauge parameter $\xi$ is found to cancel
out, and the following results are obtained for the singlet and
octet scalar contributions
\bea\lbl{scalar}
&&\correl{AAQ_AQ_A}_2=16{\tilde c^2_d} \left(1-{1\over d} \right)
\int {\iddp}
\,{1\over p^2(M^2_{S_0}-p^2) }
\nonumber\\
&&\correl{AAQ_AQ_A}_3=8{ c^2_d} \left(1-{1\over d} \right)
\int {\iddp}
\,{1\over p^2(M^2_{S}-p^2) }\ .
\ena
As before, we have replaced the term 
$p_\alpha p_\beta$ in the integrands by $p^2 g_{\alpha\beta}/d $.
There are three 
additional tadpole diagrams. Let us quote the result for completeness
\be\lbl{scalartad}
<A_\alpha^a A_\beta^b Q^c_A Q^d_A>^{tad}=ig_{\alpha\beta}\left(
{8c^2_d\over M^2_S}\, d^{sab}d^{scd} +{16\tilde c^2_d\over M^2_{S_0}}
\,\delta^{ab}\delta^{cd}\right)
\int {\iddp}\,{3\over p^2}\ .
\en
Computing the integrals  we obtain the following
results for the scalar resonance contributions which are non vanishing
\bea\lbl{cd2}
&&\correl{AAQ_AQ_A}^{scal}_2=12{\tilde c^2_d\over16\pi^2}
\left(\log {M^2_{S_0}\over\Lambda^2} +{1\over6}\right)
\nonumber\\
&&\correl{AAQ_AQ_A}^{scal}_3=6 {c^2_d\over16\pi^2}
\left(\log {M^2_{S }\over\Lambda^2} +{1\over6}\right)\ .
\ena 
One notices that no dependence upon the gauge parameter $\xi$ appears except
in some of the 
terms proportional to $F_A^2$ and $F_V^2$. This is to be expected 
because, quite generally,  
the part proportional to $\xi$ in the integral representation 
can be simplified by making use of
chiral Ward identities and expressed in terms of 2-point Green's 
functions\cite{bmdash}. 

\subsection{Comparison with the results of Baur and Urech}
Up to this point we have used exactly the same resonance Lagrangian terms 
as Baur and Urech\cite{bu} and we are therefore in a position to check
their results for the chiral parameters $K^r_i$. Making use of the relations
\rf{kif1}-\rf{kif4}, for each resonance term contribution to the correlators
\rf{basedef}\rf{basedef1} we can solve for the $K^r_i$'s which will then depend
on the two scales $\mu$ and $\Lambda$. In ref.\cite{bu} 
the values of the two scales have been taken to be equal.
Following this convention for the sake of comparison, 
and using the results derived above, 
we reproduce  the results of ref.\cite{bu} 
for the parameters $K^r_1$,...,$K^r_6$. 

We disagree, however,
on a number of contributions 
which concern $K^r_{12}$, $K^r_{13}$, $K^r_{14}$.
Contrary to BU, we find that the following quantities are vanishing.
\be
K^{G_VF_V}_{13}=K^{G_VF_V}_{14}=
K^{G^2_V}_{12}=K^{G^2_V}_{13}=K^{G^2_V}_{14}=
K^{G^2_VF_V^2}_{12}=K^{G^2_VF_V^2}_{13}=K^{G^2_VF_V^2}_{14}=0\ .
\en
In the case
of $K^r_{13}$, $K^r_{14}$ this follows from 
\cite{bmdash} where it is shown  that they can be obtained 
from 2-point correlators $<Q_VQ_V>$ and $<Q_AQ_A>$. 
Obviously, these can only pick
up contributions proportional to $F_A^2$ and $F_V^2$, which is indeed what
we recover here. 
Similarly, one can also obtain $K^r_{12}$ from a correlator
with three currents $\correl{PQ_AQ_V}$ which picks up no contribution
proportional to $G^2_V$.  
\section{Short distance constraints}\label{sec:countrms}
We now need to determine the divergence structure in the integral 
representation of eq.\rf{convol} according to QCD. One way to proceed is
to determine the necessary counterterms which are generated in QCD
with a source term as in eq.\rf{lagsrc}, where $F_\mu$ 
is  a dynamical (massive) photon.
This QED-like sector differs from ordinary QED by the fact that 
we have replaced 
the charge matrix $Q$ by a vector ($q_V$) as well as an axial-vector 
($q_A$) spurion source.
We must consider this theory as a kind of effective theory and we assume
that the divergences can be absorbed into a set of local Lagrangian terms.
We will restrict ourselves to an expansion quadratic in the spurion fields. 
This represents an important simplification since, for instance, we will only
need the free photon propagator and thereby avoid the problems of the photon 
propagator at higher order in the presence of axial couplings\cite{gross-jack}.
Another restriction is that we will only consider contributions which are of
leading order in $N_c$. Dimensional regularization together with  
the naive $\gamma_5$ prescription will be used.
  
\subsection{QCD+QED counterterms}
%
\begin{figure}[th]
\begin{center}
\includegraphics[width=10cm]{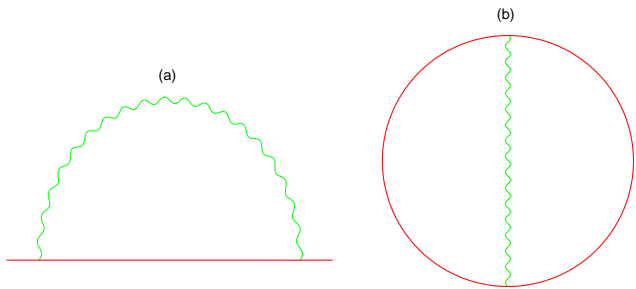}
\caption{\sl Typical divergent diagrams quadratic in the spurion sources
(a) with an open quark line (b) with a closed quark line.}
\end{center}
\end{figure}
Let us consider QCD in the presence of dynamical photons and a source term
of the form of eq.\rf{lagsrc}. The effective action is expanded to quadratic
order in the spurions $q_V$ and $q_A$ and we drop the contributions which
are subleading in $N_c$. 
We can then split the divergent terms in the effective action into two
categories, 
(a) one which collects the graphs with an open quark line and,
(b) one which collects the graphs with a closed quark loop (see fig. 1). 
The former ones are those which will prove to be of relevance to us. 

Let us first consider the contributions in category (b). 
The graphs in this category, at any order in $\alpha_s$ (in QCD the three
gluon vertex has dimension $p$, the four gluon vertex has dimension zero), 
have dimensionality
$p^4$. The corresponding local counterterms therefore must involve four 
derivatives. Examples of such terms are, 
\be
\trace{\nabla^2 q_L \nabla^2 q_L+\nabla^2 q_R\nabla^2 q_R},\ 
G_{\mu\nu} G^{\mu\nu} \trace{\nabla_\rho q_L \nabla^\rho q_L
+\nabla_\rho q_R \nabla^\rho q_R},\ \  {\rm etc...}
\en
If we restrict ourselves to the correlator combinations $\Pi_i$, eq.\rf{combi}
(from which the contribution proportional to $K_{14}$ drops out) then all the
contributions from such counterterms also drop out or are suppressed in the
large $N_c$ counting.

We are finally left with the  contributions from the first category. 
Taking into account the structure of QCD, all graphs in this category have
dimension $p$, implying that the counterterms must involve one derivative.
Furthermore, they must involve two spurions of the same kind (either $q_L$ or
$q_R$). One can form only two such independent terms
\bea\lbl{Z2}
&&{\cal L}= 
{i\over2} Z_2\,\left\{ \bar\psi_L [q_L,\nabla^\mu q_L]\gamma_\mu\psi_L 
+\bar\psi_R [q_R,\nabla^\mu q_R]\gamma_\mu\psi_R\right\}
\nonumber\\
&& +{i\over2} Z'_2\,\left\{ \bar\psi_L q_L^2 \gamma_\mu D^\mu\psi_L
+ \bar\psi_R q_R^2 \gamma_\mu D^\mu\psi_R +h.c. \right\}
\ena
where 
\bea
&&iD^\mu\psi_L= (i\partial^\mu +l_\mu + G_\mu)\psi_L,\ 
i D^\mu\psi_R= (i\partial^\mu + r_\mu + G_\mu)\psi_R\ .
\ena
At order zero in the strong coupling constant one has the well known
QED result
\be
Z_2=Z'_2= {\mu_0^{-2\epsilon}\over 16\pi^2}
(-\xi) [\Gamma(\epsilon)+\log(4\pi)] + O(\alpha_s)\ .
\en
Equality between $Z_2$ and $Z'_2$ need not hold beyond the lowest order. 

Let us now discuss the contributions proportional to $Z_2$ and $Z'_2$
to our basic correlators \rf{basedef}\rf{basedef1}. 
The first result 
is that the contribution proportional to $Z'_2$ vanishes at leading 
order in $N_c$. This can be
shown in a pedestrian way. Because of the conservation of chirality in
one fermion loop graphs, the
contributions concern only the correlators $\correl{LLQ_LQ_L}$ and 
$\correl{RRQ_RQ_R}$, it suffices to consider the former. For a given gluon
field configuration $G_\mu(x)$ we can express the functional derivatives
with respect to the sources in terms of exact fermion propagators 
$S_G(x,y)$ and we obtain the following expression
\bea\lbl{zprim2}
&&\correl{L^a_\alpha L^b_\beta Q^c_L Q^d_L}_G = {Z'_2\over 32}
\int d^4x d^4y \Big[
\nonumber\\
&&
- \delta(y)\correl{\lambda^b\lambda^a\{\lambda^c,\lambda^d\}}
<\gamma^\alpha P_L S_G(x,0)\gamma^\beta P_L S_G(0,x)>
\nonumber\\
&&-\delta(x)\correl{\lambda^a\lambda^b\{\lambda^c,\lambda^d\}}
<\gamma^\beta P_L S_G(y,0)\gamma^\alpha P_L S_G(0,y)>
\nonumber\\
&&+\correl{\lambda^a\lambda^b\{\lambda^c,\lambda^d\}}
<\gamma^\alpha P_L S_G(x,y) \gamma^\beta P_L S_G(y,0)
[i\vec{\dslash} +\Gslash ] P_L S_G(0,x)>
\nonumber\\
&&+\correl{\lambda^b\lambda^a\{\lambda^c,\lambda^d\}}
<\gamma^\beta P_L S_G(y,x) \gamma^\alpha P_L S_G(x,0)
[i\vec{\dslash} +\Gslash ] P_L S_G(0,y)>\,\Big]\ ,
\ena 
where $P_L$ is the projector $P_L=1/2(1-\gamma_5)$ and the notation
$\vec{\dslash}$ means that the derivatives are acting on the right. 
There is  another similar
contribution coming from the hermitian conjugate part of the Lagrangian. Using
the equation obeyed by the Dirac propagator
\be
[i\vec{\dslash} +\Gslash] P_L S_G(x,y)= P_R \delta^4(x-y)\ ,
\en
we indeed find that the various terms in eq.\rf{zprim2} cancel each other.
We have thus shown that the only relevant contributions from the counterterms
are those proportional to $Z_2$. 
The calculation of these is performed by computing the functional derivatives 
according to the definition of eq.\rf{basedef} and using the following 
result for the two-point correlators that appear
\be
\int d^4x {\rm e}^{ikx}< A^s_\alpha(x) A^t_\beta(0) > = 
iF_0^2\delta^{st}\left( {k_\alpha k_\beta\over k^2} - g_{\alpha\beta} \right),
\quad
\int d^4x < V^s_\alpha(x) V^t_\beta(0) >=0 \ .
\en
We find that the counterterms contribute to the following components of
our basic correlators
\bea\lbl{z2contrib}
&& < AA Q_AQ_A>^{ct}_1 = -Z_2 F_0^2
\nonumber\\
&& < AA Q_VQ_V>^{ct}_1 = -Z_2 F_0^2
\nonumber\\
&& < AV Q_AQ_V>^{ct}_1 = -{1\over2}Z_2 F_0^2
\nonumber\\
&& < AV Q_AQ_V>^{ct}_4 = -{1\over2}Z_2 F_0^2\ .
\ena
There are no contributions to the other components. 

\subsection{Summary of the $\log\Lambda^2$ dependence}
Let us consider the combinations of correlators $\Pi_i$ introduced in 
eqs.\rf{combi}\rf{combib}. 
In all these combinations the contribution from the coupling
$K^r_{14}$ drops out. In the combinations $\Pi_1$,...,$\Pi_6$ as well as
$\Pi_7$, $\Pi_9$, the contributions from the renormalization parameter $Z_2$
also drops out, as is seen from using eq.\rf{z2contrib}. As a consequence,
these quantities must be finite. More precisely, finiteness holds in the
sense of dimensional regularization which was used to establish the form of
the counterterms in the preceding section: one must compute the photon loop
integral in $d$ dimensions and then take the limit $d\to4$ which must be
finite. From these eight combinations one can then determine the 
chiral couplings $K^r_1$,...,$K^r_6$ as well as $K^r_{13}$ and $C$. 
The situation is different for $\Pi_8$ since, in that case, 
a contribution proportional to
$Z_2$ is present. The integral is not finite in general. It is finite
to zeroth order in the strong coupling $\alpha_s$ when the gauge
parameter $\xi=0$. 
This limitation affects the determination of the coupling $K^r_{12}$.

Let us  examine how these finiteness conditions  are satisfied
when the relevant correlators are computed from the minimal resonance
Lagrangian of eqs.\rf{lags4},\rf{lagav4}. Collecting the results
presented in detail in sec.\ref{sec:minreslag}, we obtain the following
dependence upon the regularization scale $\Lambda$
\bea\lbl{lambdep}
&&\Pi_1\sim -{3\over2}\,\loglb [F_0^2+F_A^2-2G_V^2     ]
\nonumber\\
&&\Pi_2\sim -{3\over2}\,\loglb [F_0^2+F_V^2-4G_VF_V+2G_V^2{\fvd\over F_0^2}    ]
\nonumber\\
&&\Pi_3\sim-6\loglb\, [ c^2_d\, ]
\nonumber\\
&&\Pi_4=0
\nonumber\\
&&\Pi_5\sim-4\loglb\, [3\tilde c^2_d-c^2_d \,]
\nonumber\\
&&\Pi_6=0
\nonumber\\
&&\Pi_7\sim  {3\over4}\loglb [ F_V^2-F_0^2-F_A^2 ]
\nonumber\\
&&\Pi_8\sim -{3\over8}\loglb [ F_V^2-F_0^2-F_A^2 +4 (G_VF_V-F_V^2) ]
\nonumber\\
&&\Pi_9\sim  3\loglb [\mvd\fvd -\mad\fad ] \ .
\ena
At this level, there is at first 
one consistency condition which is satisfied, but trivially so,
since $\Pi_4$ is identically vanishing. In the sequel (sec.\ref{sec:omegaropi})
we will consider physically relevant interactions and attempt to satisfy the
condition  in a less trivial way.  
The consistency conditions for $\Pi_7$ and 
$\Pi_9$ can be satisfied and they generate the two Weinberg
sum rule relations
\be\lbl{wsr}
\fvd=F_0^2+\fad,\quad  \mvd\fvd=\mad\fad\ .
\en
These relations will be assumed to hold from now on. Furthermore, we will
assume the following relation to hold,
\be
G_VF_V= F_0^2
\en
which follows\cite{eglpr} from imposing that the asymptotic behaviour
of the vector form-factor of the pion
\be\lbl{asyff}
\lim_{t\to\infty} t F^V_\pi(t)=0
\en
is satisfied. Using this, we observe that there are only two consistency
conditions which are fulfilled, which are those associated with
$\Pi_7$ and $\Pi_9$. The corresponding evaluation of 
$\Pi_7$ and $\Pi_9$ gives
\bea
&&\Pi_7= {3\over4}{F_0^2\over16\pi^2}\left[ \log{\mvd\over\mgd}+2 
-{\logz\over z-1} \right]
\nonumber\\
&&\Pi_9= 3\mvd {F_0^2\over16\pi^2} {\logz\over z-1}\ ,
\ena
where $z$ denotes the mass ratio 
\be
z={\mad\over\mvd}
\en
which one expects to be approximately equal to 2.
We expect $\Pi_5$ and $\Pi_6$ to vanish as a result of the leading large
$N_c$ approximation. This condition is trivially satisfied for $\Pi_6$ 
and it can be imposed for $\Pi_5$ by setting
$3\tilde c_d^2=c_d^2$. 

\section{Beyond the $O(p^4)$ resonance Lagrangian}
\subsection{Survey of $O(p^6)$ terms}
In order to comply with the other consistency conditions, it is necessary
to enlarge the resonance Lagrangian of 
eq.\rf{lagres}.  
Since this Lagrangian (for a given resonance content)
contains all terms of order $p^4$, it is natural to investigate
terms of order six. Generically, let us designate a resonance building block
by $R$ (which counts as $p^2$), 
a chiral building block by $\pi$ (which counts as $p$) 
and a source term by $J$ (which counts as $p^2$).  
In this schematic notation, there are four classes of terms of
order $p^4$ which involve at least one resonance
\be
RR,\ RJ,\ R\pi,\ R\pi\pi\ .
\en
At order $p^6$ one can have the same terms but involving two more 
derivatives and, in addition, one can have terms  formed from three,
four as well as five building blocks
\be
RRR,\ RR\pi,\ RRJ,\ RJJ,\ RJ\pi,\ RR\pi\pi,\ RJ\pi\pi,\ R\pi\pi\pi,\ 
R\pi\pi\pi\pi\ .
\en
We will not attempt to classify and discuss all 
possible remaining terms in this work. We will simply consider 
(except in sec.\ref{sec:omegaropi} ) sets of terms of the form 
$RR\pi$. The first set will be labeled as $a_1\rho\pi$ and has the
following form
\bea\lbl{lagGH}
&&{\cal L}_{a_1\rho\pi}=i(G_1 g^{\mu_0\mu_1}g^{\mu_2\mu3} +
              G_2 g^{\mu_0\mu_2}g^{\mu_3\mu_1} +
              G_3 g^{\mu_0\mu_3}g^{\mu_1\mu_2})\correl{
\,[\nabla_{\mu_0} V_{\mu_1\lambda},A_{\mu_2}^{\phantom{\mu_2}\lambda}] 
u_{\mu_3}}
\nonumber\\
&&\ +i(H_1 g^{\mu_0\mu_1}g^{\mu_2\mu_3} +
              H_2 g^{\mu_0\mu_2}g^{\mu_3\mu_1} +
              H_3 g^{\mu_0\mu_3}g^{\mu_1\mu_2})\correl{
\,[V_{\mu_1\lambda},A_{\mu_2}^{\phantom{\mu_2}\lambda}] \nabla_{\mu_0} 
u_{\mu_3}}\ .
\ena
We will first show that by 
adding these terms to the minimal resonance Lagrangian
we can reproduce the $\correl{VAP}$ 3-point function  constructed in 
ref.\cite{bmdash} 
\footnote{An extension  was recently 
considered\cite{ecker-vap} which includes a $\pi'$ resonance multiplet 
and a corresponding resonance chiral Lagrangian was constructed.
The result for the coupling constant combination $G_1+G_2+2G_3$ 
which is needed in the following turns out to be 
unaffected by this extension.}.
In addition, we will then show that it is possible
to satisfy the consistency conditions associated with $\Pi_1$ and $\Pi_2$,
and these provides a number of  Weinberg-type sum rule relations. 
In an analogous way, we introduce
an $a_0f_1\pi$ coupling (we assume nonet symmetry which reduces the number
of couplings),
\be\lbl{lagsapi}
{\cal L}_{a_0f_1\pi}=c_A\trace{S\{\nabla^\mu A_{\mu\nu}, u^\nu\} }
\en
which will allow us to satisfy the consistency condition on $\Pi_3$. 

Finally,
we will consider a set of terms of the type $\omega\rho\pi$ which contribute
to $\Pi_4$.  A calculation which is equivalent to 
an evaluation of this contribution was performed in ref.\cite{am}
using the vector, instead of the anti-symmetric tensor, formulation for the
vector fields. A classification of the terms in this latter formulation
was recently presented by Ruiz-Femenia et al.\cite{portoles}, 
we list below those of relevance to us,
\bea\lbl{omropi}
&&{\cal L}_{\omega\rho\pi}=\epsilon_{\mu\nu\rho\sigma}\,\Big\{
d_3 \trace{\{ \nabla_\lambda V^{\mu\nu},V^{\rho\lambda}\} u^\sigma}
 +d_4\trace{\{ \nabla^\sigma V^{\mu\nu},V^{\rho\lambda}\} u_\lambda}
\nonumber\\
&&\phantom{{\cal L}_{\omega\rho\pi}=\epsilon_{\mu\nu\rho\sigma}}
 +{c_5\over M_V}\trace{\{ \nabla_\lambda V^{\mu\nu},
f_+^{\rho\lambda}\} u^\sigma}
   +{c_6\over M_V}\trace{\{ \nabla_\lambda V^{\mu\lambda},
f_+^{\rho\sigma}\} u^\nu}
\nonumber\\
&& \phantom{{\cal L}_{\omega\rho\pi}=\epsilon_{\mu\nu\rho\sigma}}  
+{c_7\over M_V}\trace{\{ \nabla^\sigma V^{\mu\nu},
f_+^{\rho\lambda}\} u_\lambda}\Big\} \ .
\ena
In enabled them to reproduce the 3-point function
$\correl{VVP}$ as constructed in ref.\cite{bmwz6}. 
The consistency condition on $\Pi_4$ is somewhat more complicated  
to satisfy in the antisymmetric tensor  than in the vector formalism.  
This will be explained in more detail below.

\subsection{ ${\cal L}_{a_1\rho\pi}$ constraints from $\correl{VAP}$ }
Let us examine the constraints on the six coupling constants of the 
$a_1\rho\pi$ Lagrangian \rf{lagGH} which arise from the 3-point correlator
$\correl{VAP}$. This correlator  is defined as follows
\be
W_{\alpha\beta}(p,q)=\int d^4x d^4y {\rm e}^{ipx+iqy}\correl{0\vert T\,
(V^1_\alpha(x) A^2_\beta(y) P^3(0))\vert 0}\ .
\en
Assuming an ansatz in terms of rational functions with a minimal number
of resonance poles (i.e. one pion, one vector and one axial-vector resonance
poles) it can be shown\cite{bmdash} that 
this correlator is uniquely determined from a) the 
chiral Ward identities, b) the constraint to match the leading asymptotic
scaling behaviour,
\be
\lim_{\lambda\to\infty}W_{\alpha\beta}(\lambda p,\lambda q)=
{<\bar uu>\over
\lambda^2}\left
\{ {(p_\alpha+2q_\alpha)q_\beta-q^2g_{\alpha\beta}\over q^2 l^2}
  +{(p^2-q^2-l^2)P_{\alpha\beta}\over 2p^2 q^2 l^2} 
  -             {Q_{\alpha\beta}\over p^2 q^2 l^2}  \right\}
\en
with $l^2=(p+q)^2$ ,
\bea
&&P_{\alpha\beta}=p_\beta q_\alpha-p.q g_{\alpha\beta}\nonumber\\
&&Q_{\alpha\beta}=p^2 q_\alpha q_\beta +q^2 p_\alpha p_\beta 
-p.q p_\alpha q_\beta -p^2 q^2 g_{\alpha\beta},
\ena
and, finally, c) the asymptotic constraint on the vector form factor
of the pion eq.\rf{asyff}. More detailed analysis of asymptotic constraints
related to this 3-point correlator were performed in ref.\cite{nyffeler}. 
The result, for $\correl{VAP}$, reads
\bea
&&W_{\alpha\beta}(p,q)=<\bar uu>\Bigg\{
{ (p_\alpha+2q_\alpha)q_\beta-q^2 g_{\alpha\beta}\over q^2 l^2}
\nonumber\\
&&\phantom{W_{\alpha\beta}(p,q)= }
+{F(p^2,q^2,l^2)\,P_{\alpha\beta}\over (p^2-\mvd)(q^2-\mad)\,l^2} 
+{G(p^2,q^2,l^2)\,Q_{\alpha\beta}\over (p^2-\mvd)(q^2-\mad)q^2 l^2}\Bigg\}
\ena
with
\bea\lbl{FG}
&&F(p^2,q^2,l^2)={1\over2}(p^2-q^2-l^2)+ \mvd-\mad
\nonumber\\
&&G(p^2,q^2,l^2)= -q^2 +2\mad\ .
\ena

Let us compute the $VAP$ Green's function from the resonance Lagrangian
$\lag^{(4)}_{res}+\lag_{a_1\rho\pi}$.
We find, for the $F$ and $G$ polynomials,
\bea\lbl{FGlag}
&&F(p^2,q^2,l^2)= 
p^2\left[-{\fad\over F_0^2}-{F_VF_A\over F_0^2}(-2G_1-G_2-2G_3 +H_1+H_2 )
\right]
\nonumber\\
&&\phantom{F(p^2,q^2,l^2)}
+q^2 \left[{\fvd-2G_VF_V\over F_0^2}-{F_VF_A\over F_0^2}(G_2+2G_3 +H_1+H_2 )
\right]
\nonumber\\
&&\phantom{F(p^2,q^2,l^2)}
+l^2 \left({-F_VF_A\over F_0^2}\right)\left[-G_2-2G_3 +H_1+H_2+4H_3\right]
\nonumber\\
&&\phantom{F(p^2,q^2,l^2)}
+{1\over F_0^2}\left[\mvd \fad-\mad(\fvd-2G_VF_V)\right]
\nonumber\\
&&G(p^2,q^2,l^2)=q^2\left[ -{2G_VF_V\over F_0^2}
-{2F_VF_A\over F_0^2}(-G_1+H_1+H_2) \right]+2\mad{G_VF_V\over F_0^2}\ .
\ena
Let us first compare the constant terms in eqs.\rf{FGlag} and \rf{FG}. 
Provided that the two Weinberg relations eq.\rf{wsr} are satisfied
together with the relation $G_VF_V=F_0^2$, one finds that the constant terms 
are in exact agreement. Next, the following 
three independent relations between the coupling constants
$G_i$, $H_i$
\bea\lbl{vaprel}
&&G_1+G_2+2G_3  = {F_A\over F_V} 
\nonumber\\
&&H_1+H_2=        {F_A\over 2F_V} -{F_V\over 2F_A} +G_1
\nonumber\\
&&2H_3   =        {F_V\over 2F_A} -G_1
\ena
are found to be the necessary and sufficient conditions for the 
polynomials in eqs.\rf{FGlag} and \rf{FG} to be completely identical.

\subsection{ Linear contributions from ${\cal L}_{a_1\rho\pi}$ }
Let us now return to the 4-point correlators and consider the contributions
which are linear in the couplings of eq.\rf{lagGH}. The couplings $H_i$ do
not contribute because derivatives of the sources are involved. There
are contributions to $\correl{AAQ_AQ_A}_1$, $\correl{AAQ_VQ_V}_1$ and 
$\correl{AVQ_AQ_V}_1$ which we consider in turn. 
There is one kind of diagram relevant for $\correl{AAQ_AQ_A}_1$
\begin{center}
\begin{texdraw}
\drawdim cm 
\move(0 0)
\textref h:C v:B \htext(1.2 3.0){\qa}
\textref h:C v:B \htext(0.9 3.0){\amu}
\textref h:C v:C \htext(0.9 1.5){\amu}
\textref h:C v:C \htext(1.0 0.0){\qa}
\textref h:R v:C \htext(0.6 3.0){{$F_V,\ G_V$}}
\textref h:R v:B \htext(0.6 1.5){$G_i$}
\textref h:R v:B \htext(0.6 0.0){$F_A$}
\move(1 0) \linewd 0.20 \setgray 0.8 \lvec(1 1.5)
           \linewd 0.20 \setgray 0.5 \lvec(1 3.0)
\end{texdraw}
\end{center}
The upper vertex involves either $F_V$ or $G_V$. These diagrams give
\bea
&& \correl{AAQ_AQ_A}_1=
 -4F_A(F_V-2G_V)(1-{1\over d})
\int \iddp\Bigg[ {(d-2)G_2\over\mvd(\mad-p^2)}
\nonumber\\
&&\phantom{\correl{AAQ_AQ_A}_1= -4F_A(F_V-2G_V)(1-{1\over d}) }
+{G_{123}\over (\mad-p^2)(\mvd-p^2)}\Bigg]
\ena
where 
\be
G_{123}\equiv G_1+G_2+2G_3, 
\en
denotes the combination which was determined from $\correl{VAP}$.
In the case of the correlator $\correl{AAQ_VQ_V}_1$,
there is again one kind of diagram which makes a linear contribution

\begin{center}
\begin{texdraw}
\drawdim cm 
\move(0 0)
\textref h:C v:B \htext(1.0 3.0){\qv}
\textref h:C v:C \htext(0.9 1.5){\amu}
\textref h:C v:B \htext(0.9 0.0){\amu}
\textref h:C v:C \htext(1.2 0.0){\qv}
\textref h:R v:C \htext(0.6 3.0){$F_A$}
\textref h:R v:B \htext(0.6 1.5){$G_i$}
\textref h:R v:B \htext(0.6 0.0){$F_A$}
\move(1 0) \linewd 0.20 \setgray 0.8 \lvec(1 1.5)
           \linewd 0.20 \setgray 0.5 \lvec(1 3.0)
\end{texdraw}
\end{center}
and the result is very similar to the preceding one
\be
\correl{AAQ_VQ_V}_1 =\quad -4F_V^2(1-{1\over d})
\int \iddp\left[ {(d-2)G_1\over\mad(\mvd-p^2)}
+{G_{123}\over (\mad-p^2)(\mvd-p^2)}\right]\ .
\en
Next, in order to obtain $\correl{AVQ_AQ_V}$
one must calculate a set of five diagrams which are depicted below,
\begin{center}
\begin{texdraw}
\drawdim cm 
\move(0 0)
\textref h:C v:B \htext(0.9 3.0){\vmu}
\textref h:C v:C \htext(1.2 3.0){\qv}
\textref h:C v:B \htext(0.9 1.5){\amu}
\textref h:C v:C \htext(1.0 0.0){\qa}
\textref h:R v:C \htext(0.6 3.0){$F_A$}
\textref h:R v:B \htext(0.6 1.5){$G_i$}
\textref h:R v:B \htext(0.6 0.0){$F_V$}
\move(1 0) \linewd 0.20 \setgray 0.8 \lvec(1 1.5)
           \linewd 0.20 \setgray 0.5 \lvec(1 3.0)
\textref h:C v:B \htext(4.0 3.0){\qv}
\textref h:C v:C \htext(4.0 2.25){\vmu}
\textref h:C v:B \htext(3.9 1.5){\amu}
\textref h:C v:C \htext(4.0 0.0){\qa}
\move(4 0) \linewd 0.20 \setgray 0.8 \lvec(4 1.5)
           \linewd 0.20 \setgray 0.5 \lvec(4 3.0)
\textref h:C v:B \htext(7.0 3.0){\qv}
\textref h:C v:C \htext(6.9 1.5){\amu}
\textref h:C v:B \htext(7.2 1.5){\vmu}
\textref h:C v:C \htext(7.0 0.0){\qa}
\move(7 0) \linewd 0.20 \setgray 0.8 \lvec(7 1.5)
           \linewd 0.20 \setgray 0.5 \lvec(7 3.0)
\textref h:C v:B \htext(10.0 3.0){\qv}
\textref h:C v:B \htext( 9.9 1.5){\amu}
\textref h:C v:C \htext( 9.9 0.75){\vmu}
\textref h:C v:C \htext(10.0 0.0){\qa}
\move(10 0) \linewd 0.20 \setgray 0.8 \lvec(10 1.5)
           \linewd 0.20 \setgray 0.5 \lvec(10 3.0)
\textref h:C v:B \htext(13.0 3.0){\qv}
\textref h:C v:B \htext(12.9 1.5){\amu}
\textref h:C v:C \htext(12.9 0.0){\vmu}
\textref h:C v:C \htext(13.2 0.0){\qa}
\move(13 0) \linewd 0.20 \setgray 0.8 \lvec(13 1.5)
           \linewd 0.20 \setgray 0.5 \lvec(13 3.0)
\end{texdraw}
\end{center}
Computing these diagrams gives
\bea
&&\correl{AVQ_AQ_V}_1=-F_AF_V G_{123}(1-{1\over d})
\int \iddp\Bigg[ 
{d-2\over (\mad-p^2)(\mvd-p^2)} 
\nonumber\\
&&\phantom{\correl{AVQ_AQ_V}_1}
+{2\mad\over (\mad-p^2)^2(\mvd-p^2)}
+           {2\mvd\over (\mad-p^2)(\mvd-p^2)^2}\Bigg]
\ena
which is seen to be proportional to the coupling constant combination
$G_{123}$.

\subsection{Quadratic contributions from ${\cal L}_{a_1\rho\pi}$
and consistency conditions for $\Pi_1$, $\Pi_2$, $\Pi_8$}\label{sec:quadct}
We turn now to the contributions which are quadratic in the couplings
$G_i$. These concern only the two correlators 
$\correl{AAQ_AQ_A}_1$, $\correl{AAQ_VQ_V}_1$ and they are given respectively 
from the two kinds of diagrams depicted below
\begin{center}
\begin{texdraw}
\drawdim cm 
\move(0 0)
\textref h:C v:B \htext(1.0 3.0){\qa}
\textref h:C v:C \htext(1.0 2.0){\amu}
\textref h:C v:B \htext(1.0 1.0){\amu}
\textref h:C v:C \htext(1.0 0.0){\qa}
\textref h:R v:C \htext(0.6 3.0){$F_A$}
\textref h:R v:B \htext(0.6 2.0){$G_i$}
\textref h:R v:B \htext(0.6 1.0){$G_j$}
\textref h:R v:B \htext(0.6 0.0){$F_A$}

\move(1 0) \linewd 0.20 \setgray 0.8 \lvec(1 1.0)
           \linewd 0.20 \setgray 0.5 \lvec(1 2.0)
           \linewd 0.20 \setgray 0.8 \lvec(1 3.0)
\textref h:C v:B \htext(4.0 3.0){\qv}
\textref h:C v:C \htext(4.0 2.0){\amu}
\textref h:C v:B \htext(4.0 1.0){\amu}
\textref h:C v:C \htext(4.0 0.0){\qv}
\textref h:R v:C \htext(3.6 3.0){$F_V$}
\textref h:R v:B \htext(3.6 2.0){$G_i$}
\textref h:R v:B \htext(3.6 1.0){$G_j$}
\textref h:R v:B \htext(3.6 0.0){$F_V$}
\move(4 0) \linewd 0.20 \setgray 0.5 \lvec(4 1.0)
           \linewd 0.20 \setgray 0.8 \lvec(4 2.0)
           \linewd 0.20 \setgray 0.5 \lvec(4 3.0)
\end{texdraw}
\end{center}
The computation gives the following result for 
$\correl{AAQ_AQ_A}_1$,
\be
\correl{AAQ_AQ_A}_1=4F_A^2 (1-{1\over d})
 \int\iddp\Bigg\{ {(G_{123})^2\,p^2\over (\mad-p^2)^2 (\mvd-p^2)}
+{G_2^2\, (2-d)\,p^2\over \mvd (\mad-p^2)^2 }\Bigg\}\ .
\en
The result for $\correl{AAQ_VQ_V}_1$ is obtained simply by
replacing $F_A$ by $F_V$ and by interchanging  $M_A$, $M_V$ and $G_1$, $G_2$
in the expression above.

We can now collect all the contributions which involve the coupling constants
$G_i$,
\bea
&&\correl{AAQ_AQ_A}_1^{G_i}=
{3F_A(F_V-2G_V)\over16\pi^2}\Bigg\{ {-2G_2 \,z}
\Bigg( {7\over6}+\logA \Bigg)
\nonumber\\
&&\phantom{\correl{AAQ_AQ_A}_1} 
+G_{123}
\Bigg[ {1\over6}+ {z\logz\over z-1}+\logV \Bigg] \Bigg\}
+{3F_A^2\over16\pi^2}\,\Bigg\{ 4 G_2^2 \,z \Bigg( {5\over3}+\logA \Bigg)
\nonumber\\
&& \phantom{\correl{AAQ_AQ_A}_1}
-(G_{123})^2\Bigg[ {7z-1\over 6(z-1)}-{\logz\over (z-1)^2}
+\logA\Bigg]\Bigg\}
\\
&&\correl{AAQ_VQ_V}_1^{G_i}=
{3F_AF_V\over16\pi^2}\Bigg\{ {-2\,G_1\over z}
\left({7\over6}+\logV\right)
\nonumber\\
&&\phantom{\correl{AAQ_VQ_V}_1}
+G_{123}\Bigg[{1\over6}+{z\logz\over z-1}+\logV \Bigg]
\Bigg\}
+{3F_V^2\over16\pi^2}\Bigg\{ 4{G_1^2\over z}\Bigg({5\over3}+\logV\Bigg)
\nonumber\\
&&\phantom{\correl{AAQ_VQ_V}_1}
-(G_{123})^2\Bigg[ {7-z\over6(1-z)} +{z^2\logz\over (z-1)^2}+\logV
\Bigg]\Bigg\}\ ,
\\
&&\correl{AVQ_AQ_V}_1^{G_i}={3\over2}{F_AF_V G_{123}\over 16\pi^2}\Bigg[
{1\over6}+{ z\logz\over z-1}+ \logV\Bigg]\ .
\ena
Adding these contributions to those computed in sec.\ref{sec:minreslag}
it is now possible to satisfy the consistency conditions
associated with $\Pi_1$, $\Pi_2$ and $\Pi_8$. Firstly one
observes that the $\Lambda$ scale dependence drops out from the combination
$\Pi_8$ automatically provided the first relation \rf{vaprel} holds
for $G_{123}$. 
Requiring that the consistency conditions for $\Pi_1$ and $\Pi_2$ be
satisfied yields two Weinberg-type relations for $G_1$ and $G_2$
\bea\lbl{eqG1G2}
&&8G_1^2-{4\over\sqrt{z}}     G_1  -1 =0
\nonumber\\
&&8G_2^2 -{4\over\sqrt{z}}(2-z)G_2  -1 =0\ .
\ena
The previous determinations of $F_V$, $F_A$, $G_V$ and of the
combination $G_1+G_2+2G_3$ have been used. 
Since these equations are quadratic, one could be concerned about 
the existence of real solutions. Real solutions do exist for any value
of the mass ratio $z$ and, a priori, one has a set of four different 
solutions. By studying the behaviour as a function of $z$ we can reduce
the multiplicity. We note that the Weinberg equations \rf{wsr} have 
real solutions provided $z$ lies in the range $1\le z < \infty$. 
Firstly, let $z\to1$: this corresponds to
a situation where chiral symmetry gets restored
and one must have $G_1=G_2$ in this case. 
This argument reduces the multiplicity to
just two solutions, which we can write,
\bea
&&G_1= {1\over4\sqrt z}(1+\sigma\sqrt{1+2z})\nonumber
\nonumber\\
&&G_2= {1\over4\sqrt z}(2-z +\sigma\sqrt{(2-z)^2+2z}),\quad \sigma=\pm1\ .
\ena
We can further reduce the multiplicity by observing the behaviour as 
$z\to\infty$. In this limit, $G_1$ remains finite while in the
case of $G_2$ only one solution remains finite  corresponding to
$\sigma=1$. Correspondingly, the divergence in $\Pi_1$ is logarithmic
for $\sigma=1$ while it is linear for  $\sigma=-1$.
The solution with $\sigma=1$, therefore, appears to be the most 
plausible choice. Using this solution, the
following determinations are obtained for $\Pi_1$, $\Pi_2$ and $\Pi_8$
\bea\lbl{consist128}
&&\Pi_1={3\over2}{F_0^2\over 16\pi^2}\Bigg\{ \logg 
 +{ (z^3-5z^2+5z+1)\logz\over(z-1)^3}
\nonumber\\
&&\phantom{\Pi_1={3\over2}{F_0^2\over 16\pi^2}\Bigg\{} 
+ {z^3-2z^2+7z-10\over2(z-1)^2} - {(z-2)\sqrt{z^2-2z+4}\over2(z-1)}\Bigg\}
\nonumber\\
&&\Pi_2={3\over2}{F_0^2\over 16\pi^2}\Bigg\{ \logg 
 -{2z^2\logz\over(z-1)^3} 
\nonumber\\
&& \phantom{\Pi_1={3\over2}{F_0^2\over 16\pi^2}\Bigg\{} 
+ {6z^3-7z^2+6z-1\over2z(z-1)^2} + {\sqrt{2z+1}\over2z(z-1)}\Bigg\}
\nonumber\\
&&\Pi_8={3\over8}{F_0^2\over 16\pi^2}
\left(-\logg +{(3 z+ 1)\logz \over (z-1)^2} - {4 z\over z-1}  \right)\ .
\ena

\subsection{Contributions from ${\cal L}_{a_0f_1\pi}$ and the consistency
condition for $\Pi_3$}
By analogy with the preceding we now consider  couplings which involve the
scalar mesons of the type $a_0f_1\pi$ which will contribute to 
$\correl{AAQ_AQ_A}_{2,3}$. There is only one Lagrangian term which matters,
shown in eq.\rf{lagsapi} (assuming nonet symmetry). 
The term in which the derivative acts on $u_\nu$ does not contribute
to our correlators nor to the physical decay amplitude which is
discussed below. 
There are two kinds of diagrams which contribute to the relevant 
correlators
\begin{center}
\begin{texdraw}
\drawdim cm 
\textref h:C v:C \htext(1.1 3.0){\qa}
\textref h:C v:C \htext(0.9 3.0){\amu}
\textref h:C v:C \htext(0.9 1.5){\amu}
\textref h:C v:C \htext(1.0 0.0){\qa}
\textref h:R v:C \htext(0.6 3.0){$c_d$}
\textref h:R v:B \htext(0.6 1.5){$c_A$}
\textref h:R v:B \htext(0.6 0.0){$F_A$}
\move(1 0) \linewd 0.20 \setgray 0.8 \lvec(1 1.5)
           \linewd 0.20 \setgray 0.2 \lvec(1 3)
\textref h:C v:C \htext(4.0 3.0){\qa}
\textref h:C v:C \htext(3.9 2.0){\amu}
\textref h:C v:C \htext(3.9 1.0){\amu}
\textref h:C v:C \htext(4.0 0.0){\qa}
\textref h:R v:C \htext(3.6 3.0){$F_A$}
\textref h:R v:B \htext(3.6 2.0){$c_A$}
\textref h:R v:B \htext(3.6 1.0){$c_A$}
\textref h:R v:B \htext(3.6 0.0){$F_A$}
\move(4 0) \linewd 0.20 \setgray 0.8 \lvec(4 1)
           \linewd 0.20 \setgray 0.2 \lvec(4 2)
           \linewd 0.20 \setgray 0.8 \lvec(4 3)
\end{texdraw}
\end{center}
Computing the diagrams gives
\bea
&&\correl{AAQ_AQ_A}_3=\left(1-{1\over d}\right)\int
\iddp \Bigg\{ {8\sqrt{2}c_dc_AF_A\over (\mad-p^2)(\msd-p^2)}
\nonumber\\
&&\phantom{\correl{AAQ_AQ_A}_3}
+{4c_A^2\fad\,p^2\over (\mad-p^2)^2(\msd-p^2)}\Bigg\}
\nonumber\\
&&\phantom{\correl{AAQ_AQ_A}_3}
={-6\sqrt{2}c_Ac_dF_A\over16\pi^2}\left({1\over6}+\logS+
{\mad\over\mad-\msd}\log{\mad\over\msd}\right)
\\
&&\phantom{\correl{AAQ_AQ_A}_3}
+{3c_A^2\fad\over16\pi^2}\left({1\over6}+\logS+
{\mad\over\mad-\msd}+ {\mad(\mad-2\msd)\over(\mad-\msd)^2}
\log{\mad\over\msd}\right)\ .
\nonumber
\ena
We can now add this result to the contribution \rf{cd2} 
proportional to $c_d^2$, and the
consistency condition translates into a simple relation between the coupling
constants $c_d$ and $c_a$
\be\lbl{cAval}
c_AF_A=\sqrt{2}c_d\ .
\en
This allows one to obtain a consistent determination of the correlator
component\  $\correl{AAQ_AQ_A}_3$ associated with the scalar mesons,
\be
\Pi_3=\correl{AAQ_AQ_A}_3={6c^2_d \mad\over16\pi^2 (\mad-\msd)}\left(
1- {\mad\over\mad-\msd}\log{\mad\over\msd}               \right)\ .
\en 

We can compare the above determination of the coupling $c_A$ with experiment.
Let us consider the decay process $f_1(1285)\to a_0(980)+\pi$.
Using the Lagrangian above \rf{lagsapi} the expression for the decay width
is easily derived,
\be
\Gamma_{f_1\to a_0\pi}= 
c_A^2 {p_{cm}\,(\mad-\msd)^2 \over8\pi M^2_{f_1} F_0^2}\ .
\en
Using the value $c_d\simeq 32$ MeV from ref.\cite{eglpr}, $M_A=\sqrt{2} M_\rho$
(and physical masses in kinematical factors) one obtains 
\be
\Gamma_{f_1\to a_0\pi}\simeq 7.7\ {\rm MeV}\ .
\en
This result is in rather good agreement with the experimental one 
$\Gamma^{exp}_{f_1\to a_0\pi}= 8.6\pm 1.7$ MeV 
from ref.\cite{pdg}.

\subsection{Contributions from ${\cal L}_{\omega\rho\pi}$ 
in the vector v/s tensor formalism and $\Pi_4$}\label{sec:omegaropi}
We discuss here contributions from ${\cal L}_{\omega\rho\pi}$ eq.\rf{omropi}
which are likely to be important and were also not accounted for in the minimal
resonance Lagrangian. One faces a technical difficulty  that the
Lagrangian eq.\rf{omropi} involves the Levi-Civitta $\epsilon$ tensor and this
makes continuation to $d$ dimensional space ambiguous. One would therefore
like to impose a stronger version of the consistency condition that the
integral be finite in four dimensions. This condition is rather
simple to enforce if one uses the vector formalism. Let us first briefly
review what would be the result in this case  and then
we will show how to proceed in the case of the tensor formalism.
An extensive study of resonance Lagrangian terms in the vector
formalism can be found in ref.\cite{prades}.
In this formulation, the terms 
including the $\omega\rho\pi$ type couplings read (we use the
same notation as \cite{eglpr}),
\bea\lbl{vecform}
&& {\cal L}_{\omega\rho\pi}^{V}=
-{1\over4}\trace{ V_{\mu\nu} V^{\mu\nu} 
-2 m^2_V V_\mu V^\mu}
-{1\over2\sqrt2}f_V\trace{V_{\mu\nu} f_+^{\mu\nu}}
\nonumber\\
&& +{g^V_1\over2\sqrt2}\,\eeps \trace{ \{\um, V_\nu\}\fpab }
   +{g^V_2\over2      }\,\eeps \trace{ \{\um, V_\nu\} V_{\alpha\beta }}\ ,
\ena
where
\be
V_{\mu\nu}=\nabla_\mu V_\nu -\nabla_\nu V_\mu\ .
\en
In this description the consideration of short distance conditions imposes
that chiral $O(p^4)$ terms be added\cite{eglpr} but these do not play any
role for our calculation.
Computing the correlator $<AAQ_VQ_V>_3$ from this Lagrangian, one finds
\be\lbl{epsvec}
<A_\alpha A_\beta Q_VQ_V>_3=8 \int { d^4p\over (2\pi)^4}{
(p^2g_{\alpha\beta}-p_\alpha p_\beta)
\left[ p^2(g^V_1+2g^V_2f_V)-\mvd g^V_1 \right]^2\over p^2 (\mvd -p^2)^3 }
\en
The integral in eq.\rf{epsvec} can be defined in four dimensions without
difficulty, it suffices  that the following relation holds among
the coupling constants
\be
g^V_1+2g^V_2f_V=0\ .
\en
The result for the correlator then reads,
\be\lbl{gvres}
<AAQ_VQ_V>^{\omega\rho\pi}_3={F_0^2\over 16\pi^2}(3\tilde g_V^2),\quad 
\tilde g_V={g_1^V M_V\over F_0}\ .
\en
The value of $\tilde g_V$ was estimated in ref.\cite{am} from the 
$\omega\to\pi\gamma$ decay rate, $\vert\tilde g_V\vert=0.91\pm  0.03 $.
Under the assumption of nonet symmetry (which is rather well satisfied
experimentally for the vectors), one also has
\be\lbl{gvres1}
<AAQ_VQ_V>^{\omega\rho\pi}_2={2\over3}<AAQ_VQ_V>^{\omega\rho\pi}_3\ .
\en
From eqs.\rf{gvres},\rf{gvres1} one can deduce an estimate, which is identical
to the one performed in ref.\cite{am}, for the 
combination of $K_i$'s which is involved in the radiative corrections
to the process $\pi^0\to2\gamma$. 

Let us now reconsider the problem using the antisymmetric tensor formulation,
i.e. starting from the Lagrangian of eq.\rf{omropi}. 
One has three kinds of diagrams to compute which are shown below
\begin{center}
\begin{texdraw}
\drawdim cm 
\move(0 0)
\textref h:C v:C \htext(0.9 3.0){\qv}
\textref h:C v:C \htext(1.2 3.0){\amu}
\textref h:C v:C \htext(1.2 0.0){\amu}
\textref h:C v:C \htext(0.9 0.0){\qv}
\textref h:R v:C \htext(0.6 3.0){$c_i$}
\textref h:R v:C \htext(0.6 0.0){$c_j$}
\linewd 0.20
\setgray 0.5
\move(1.0 0.0)  \lvec(1 3)
\move(4 0)
\textref h:C v:C \htext(4.0 3.0){\qv}
\textref h:C v:C \htext(4.0 1.5){\amu}
\textref h:C v:C \htext(3.9 0.0){\amu}
\textref h:C v:C \htext(4.2 0.0){\qv}
\textref h:R v:C \htext(3.6 3.0){$F_V$}
\textref h:R v:C \htext(3.6 1.5){$d_i$}
\textref h:R v:C \htext(3.6 0.0){$c_j$}
\linewd 0.20
\setgray 0.5
\move(4.0 0.0)  \lvec(4 3)
\move(7 0)
\textref h:C v:C \htext(7.0 3.0){\qv}
\textref h:C v:C \htext(6.9 2.0){\amu}
\textref h:C v:C \htext(6.9 1.0){\amu}
\textref h:C v:C \htext(7.0 0.0){\qv}
\textref h:R v:C \htext(6.6 3.0){$F_V$}
\textref h:R v:C \htext(6.6 2.0){$d_i$}
\textref h:R v:C \htext(6.6 1.0){$d_j$}
\textref h:R v:C \htext(6.6 0.0){$F_V$}
\linewd 0.20
\setgray 0.5
\move(7 0) \lvec (7 3)
\end{texdraw}
\end{center}
Evaluating these various diagrams 
we obtain  $<A_\alpha A_\beta Q_VQ_V>_3$ in the form of the following 
four-dimensional integral
\bea\lbl{vvpi}
&&\correl{A_\alpha A_\beta Q_VQ_V}_3=
\int  {d^4p\over (2\pi)^4}\,\Bigg\{ {48\,p_\alpha p_\beta \over M^4_V}
\left( {M_VF_V (d_3+d_4)\over \mvd-p^2} -\sqrt2 (c_5+c_7)\right)^2 \\
&&\phantom{\correl{A_\alpha A_\beta}_3}
+
{32(p^2 g_{\alpha\beta}-p_\alpha p_\beta)\over\mvd( \mvd-p^2)}
\left({2M_VF_V d_3\over \mvd-p^2}
-  \sqrt2 (c_5-c_6)\right)^2\Bigg\}\  .\nonumber
\ena
The integral in eq.\rf{vvpi} 
does not converge in four dimensions unless all the couplings $c_i$ and $d_i$
vanish identically. 
In order to remove the divergence, in this case,
we introduce a more general set of resonance Lagrangian  terms,
which include four building blocks
\bea\lbl{lagcont}
&&{\cal L}^{4bb}=(x_1 g^{\mu_0\mu_1} g^{\mu_2\mu_3} 
+x_2 g^{\mu_0\mu_2} g^{\mu_1\mu_3} +x_3 g^{\mu_0\mu_3} g^{\mu_1\mu_2})
\trace{\{ f^{+\lambda}_{\phantom{+\lambda}\mu_0}, u_{\mu_1}\}
\{f^+_{\lambda\mu_2}, u_{\mu_3}\} } 
\nonumber\\
&&\quad +\sqrt2(y_1 g^{\mu_0\mu_1} g^{\mu_2\mu_3} 
+y_2 g^{\mu_0\mu_2} g^{\mu_1\mu_3} +y_3 g^{\mu_0\mu_3} g^{\mu_1\mu_2})
\trace{\{ V^{\lambda}_{\phantom{\lambda}\mu_0}, u_{\mu_1}\}
\{f^+_{\lambda\mu_2}, u_{\mu_3}\} } 
\nonumber\\
&&\quad +(z_1 g^{\mu_0\mu_1} g^{\mu_2\mu_3} 
+z_2 g^{\mu_0\mu_2} g^{\mu_1\mu_3} +z_3 g^{\mu_0\mu_3} g^{\mu_1\mu_2})
\trace{\{ V^{\lambda}_{\phantom{\lambda}\mu_0}, u_{\mu_1}\}
\{V_{\lambda\mu_2}, u_{\mu_3}\} }\ .
\ena
These terms generate the following additional contribution to the correlator
\bea\lbl{corrcont}
&&\trace{A_\alpha A_\beta Q_VQ_V}^{4bb}_3=
\int {d^4p\over (2\pi)^4}
\Bigg\{ 24p^2 g_{\alpha\beta}\left( {2x_{123}\over p^2}
-{F_V y_{123}\over p^2(\mvd-p^2)}+{F_V^2 z_{123}\over p^2(\mvd-p^2)^2}\right)
\nonumber\\
&&\phantom{\correl{A_\alpha A_\beta }   } 
-16(p^2 g_{\alpha\beta}-p_\alpha p_\beta)
\left( {2x_{13}\over p^2}
-{F_V y_{13}\over p^2(\mvd-p^2)}+{F_V^2 z_{13}\over p^2(\mvd-p^2)^2}\right)
\Bigg\}\ .
\ena
In this expression we have used the  notation
\be
x_{123}=x_1+2x_2+x_3,\quad  x_{13}=x_1+x_3\ ,
\en
and similarly with $y_i$, $z_i$. 

The convergence constraint  
in four dimension can now be satisfied:
it imposes $c_5+c_7=0$ and determines the values of the six 
combinations of parameters which appear in eq.\rf{corrcont} in terms of
$d_3$, $d_3+d_4$ and $c_5-c_6$. Collecting all the pieces, one finds
that  all the couplings drop out except $d_3$ and the correlator 
$\correl{A_\alpha A_\beta Q_VQ_V}_3$ is then expressed as a convergent
integral,
\be
\correl{A_\alpha A_\beta Q_V Q_V}_3=128 \mvd\fvd d_3^2\int {d^4p\over (2\pi)^4}
{p^2g_{\alpha\beta}-p_\alpha p_\beta\over p^2 (\mvd-p^2)^3 }\ .
\en 
This expression is exactly analogous to the one obtained upon using the
vector formalism and we write the final result in the same form,
\be
\Pi_4=\correl{AAQ_VQ_V}_3= {F_0^2\over 16\pi^2} (3\tilde g_{AT}^2),
\quad \tilde g_{AT}={4F_Vd_3/ F_0}\ .
\en
The value of $d_3$ may be determined
by enforcing asymptotic conditions concerning the $\correl{VVP}$ 
correlator. Not all conditions can be satisfied at the same time
if one uses a minimal resonance model\cite{bmwz6,nyffeler,bij3pt}.
If one chooses to enforce the overall scaling behaviour one 
obtains\cite{portoles}
\be
d_3^{scaling}= {F_0^2\over8\mvd}-{N_c\mvd\over 64\pi^2\fvd}\ .
\en
Numerically, this gives $\tilde g_{AT}=-0.58$ which is somewhat smaller
in magnitude than $\tilde g_V$. 
Alternatively,  one may choose instead to impose that
the $\correl{VVP}$ correlator satisfies the VMD property, in which
case one gets 
\be
d_3^{VMD}= -{N_c\mvd\over 64\pi^2\fvd}\ .
\en
If one uses this latter determination, one gets $\tilde g_{AT}\simeq -0.93$, 
i.e. the result from the antisymmetric tensor formalism becomes essentially 
identical to the one from the vector formalism.

\section{Numerical results}
\subsection{Results for $K_1,...,K_6$}
\begin{table}[hbt]
\begin{center}
\begin{tabular}{|c|c|c|c|c|c|l|}\hline\hline
$10^3\,K^r_1$ &$10^3\,K^r_2$ &$10^3\,K^r_3$ &$10^3\,K^r_4$ 
&$10^3\,K^r_5$ &$10^3\,K^r_6$ &\  \\ 
\hline
-4.44 & -1.73 & 4.44 & -3.46 & 13.31 & 5.19 & $\rho,\ a_1$ \\
-6.8  & -2.7  & 6.8  & -5.5  & 20.3  &  8.2 & $\rho,\ a_1$ (BU ) \\
\hline
 1.97 &  1.97 &-1.97 &  3.93 & -1.97 &-1.97 & $\omega\rho\pi$ \\
\hline
-0.35 &  0.35 & 0.35 &  0.69 & 0.35  &-0.35 &  scalars      \\
 0.4   &  -0.4 & -0.4 & -0.7  & -0.4  & 0.4  & scalars (BU )\\ 
\hline
0.11 & 0.11 & -0.11 &  0.21 & -0.11 & -0.11 & $b_1(1235)$ \\
\hline
-2.71 & 0.69 & 2.71 & 1.38 & 11.59 & 2.77 & Total\\ \hline\hline
\end{tabular}
\caption{\sl Numerical results obtained for $K_i^r(\mu)$ with $\mu=0.77$ GeV
from the present work. The contributions associated with the $b_1(1235)$ 
are discussed in appendix B.
The results  from ref.\cite{bu} are also shown for comparison.     }
\lbltab{kivals}
\end{center}
\end{table}
In the numerical applications we will simply use $M_V=0.77$ GeV and
$F_V=\sqrt2 F_\pi$ (KSRF relation\cite{ksrf}), which corresponds to $z=2$
(the same values were used by BU). 
We will also use $F_0=F_\pi=92.4$ MeV.
The results for $K^r_1(\mu),...,
K^r_6(\mu)$ with $\mu=M_V$ are
collected in Table 1. We show separately the contributions due to the
scalars (terms proportional to $c_d$, with $c_d=32$ MeV) 
and the contributions from the
$\omega\rho\pi$ interactions (terms proportional to $\tilde g_{AT}$, with
$\vert\tilde  g_{AT}\vert =0.93$). 
We find that this contribution can be significant for some of the couplings
like $K^r_2$ or $K_4^r$.
The contributions related to the scalars are 
rather small. In that case, our results agree on the magnitude but have 
a different sign from those of BU because of the extra contribution
needed to ensure consistency.
The largest contributions are those originating 
from the $a_1\rho\pi$ sector.  In that case our 
numerical results are in qualitative agreement with those of BU.
We have also performed an estimate of the contributions associated
with the $b_1$ meson (see appendix B) which turn out to be rather small. 

\subsection{An update on the corrections to Dashen's theorem}
A reliable evaluation of the electromagnetic part of the $K^+-K^0$
mass difference is important for the determination of $m_u-m_d$. At leading
order in the chiral expansion it is given by Dashen's theorem\cite{dashen}
\be
( \Delta M^2_K )_{EM}= ( \Delta M^2_\pi )_{EM} +O(e^2 m_q )\ .
\en
It was  pointed out that the corrections of order $e^2m_s$
could be rather large\cite{maltman} 
such that the right hand side could be modified by as 
much as a factor of two. Using Urech's formalism one can compute explicitly
these corrections which have the following form\cite{urech}
\be
( \Delta M^2_K )_{EM}- ( \Delta M^2_\pi )_{EM} =e^2 M^2_K\, (A_1+A_2+A_3)
+O(e^2 M^2_\pi)
\en
with
\bea
&& A_1=-{1\over16\pi^2}[ (3+2Z)\log{M^2_K\over\mu^2}-4]-16 Z L^r_5,\quad
Z={C\over F_0^4}\ 
\nonumber\\
&& A_2=-{4\over3}( K^r_5+K^r_6)\nonumber\\
&& A_3= 8(K^r_{10}+K^r_{11} )\ .
\ena
As  discussed in (I) (and also in \cite{bij-prad}) 
the Urech couplings $K^r_{10}$, $K^r_{11}$ taken 
separately suffer from short distance ambiguities and depend on the gauge.
The sum, however, is perfectly 
well defined and can be estimated from   sum rules
related to the correlators $\correl{V^3V^3-V^8V^8-A^3A^3+A^8A^8}$ and 
$\correl{VAP}$, giving the following result\cite{bmdash}
\bea\lbl{K10pK11}
&&K^r_{10}+K^r_{11}=
{3\over16}{1\over16\pi^2}\Bigg\{ \log{M_V^2\over\mu^2}
+Z_A-Z_V\nonumber\\
&&\phantom{K^r_{10}+K^r_{11}=}
+2\logz\left[ {(z+1)^2\over2(z-1)^3}+1\right] +{2z(z-3)\over(z-1)^2}
+{1\over 6} +4Z^0_\mu  \Bigg\}
\ena
with
\be
Z_A-Z_V= {4\mad(F_\pi-F_K)\over F_\pi(\mpid-\mkd)}
+{2\fvd M_\rho(M_\rho-M_\phi)\over F_\pi^2(\mpid-\mkd)}\logz -2\logz-2\ .
\en
The quantity $Z_A-Z_V$ is generated from flavour symmetry breaking at first
order in the vector multiplet as well as the axial-vector multiplet and the
computation takes into account Weinberg-type sum rule relations for the
splittings of the masses as well as the coupling constants.
The last entry in the expression \rf{K10pK11}, $Z^0_\mu$, 
is a contribution which should restore the scale dependence proportional
to $C/F_0^4$. Since this effect 
is subleading in $N_c$, this term  will be dropped here for consistency
with previous approximations. Making use of the result eq.\rf{K10pK11} 
and of the numerical values for $K^r_5+K^r_6$ shown in table 1 we find the
following numerical values for the three corrective terms
(which we give in units of the physical $\pi^+-\pi^0$ mass difference,
$\Delta M^2_\pi =1261.16\ {\rm MeV}^2$ )
\be
( \Delta M^2_K )_{EM}- ( \Delta M^2_\pi )_{EM} = \Delta M^2_\pi\,[
0.59 -0.34  + 1.25 ] \ .
\en
The new result with respect to (I) concerns $K_5+K_6$ and we find that
this contribution goes in the sense of reducing the size of the correction.
This is in qualitative agreement with the result of BU. However, the
effect of the extra  terms from ${\cal L}_{a_1\rho\pi}$ included 
here is to reduce
the magnitude of $K_5+K_6$ by approximately a factor of two as compared
to BU. There is a clear 
indication for a sizable correction to Dashen's theorem rather similar
to that found in refs.\cite{dono-perez,bij-prad}. The origin of
the different result obtain in ref.\cite{bu} can be traced, essentially,  
to the fact that their result for $K_{10}+K_{11}$ is ten times smaller 
than ours. 
This is due to two reasons 1) flavour symmetry breaking 
in the vector and the axial-vector multiplets was ignored in \cite{bu} 
such that $K_{10}$ was just set to zero and 2) the effect of the 
couplings from ${\cal L}_{a_1\rho\pi}$ which they did not take into
account is to strongly enhance the value of $K_{11}$.
\subsection{Uncertainties}
One should keep in mind that there are  uncertainties
which are difficult to estimate quantitatively. These have two main sources.
The first one is that we have used, at several places, large $N_c$
approximations a) in the treatment of the resonances, which are taken
to be infinitely narrow and b) in the calculation of QCD n-point correlators
in which only tree graphs were considered. The second source of uncertainty
is that we have retained the contributions from the lightest resonances only. 
The usual experience with such approximations is that reasonable order of
magnitudes should be obtained.

\section{Conclusions}
In this paper, we have considered a set of sum rules for the electromagnetic
chiral parameters which involve QCD 4-point correlators in connection
with the construction of a chiral Lagrangian with resonances. 
For our purposes, the resonance Lagrangian provides rational-type  
approximations for the QCD correlators obeying the chiral Ward identities.

We have shown that the resonance Lagrangian of order $p^4$ 
was not sufficient to ensure all the convergence conditions 
required by the sum rules.
We have then shown that introducing a set of terms of order  $p^6$ 
involving $a_1\rho\pi$--type couplings renders it possible to satisfy 
these conditions in the sense of finiteness as a $d\to4$ limit.  
The new coupling constants which appear were shown to 
obey non linear equations, we have argued that physical requirements 
select a unique solution.

Our investigation, however, was essentially 
limited to a set of terms with three
building blocks. A more general investigation should be performed
in the future. 
A more complete determination of the resonance Lagrangian would 
lead to much better estimates of the chiral coupling constants
at this order\cite{bcep6} than those available at present. This would improve
significantly the effectiveness of the chiral calculations at order $p^6$.
There is also a limitation in the set of resonances
which were included. In this regard, the role of tensor mesons should 
perhaps be investigated\cite{toublan}.

We have considered an application to 
the problem of the chiral corrections to Dashen's low-energy theorem. 
We have found that the combination $K^r_5+K^r_6$ as determined here
reduces the size of the correction which still
remains rather large. 
One motivation for deriving estimates for each of the parameters $K_i$
is to improve the accuracy of radiative correction evaluations. In this 
context, further work is still needed because another set of parameters, 
$X_i$, is involved in the interesting 
case of the semi-leptonic decays\cite{knecht-lept}. 
Finally, a  formal similarity can be noted between the sum rules 
considered here
and those shown recently to hold for chiral parameters in the weak non-leptonic
sector\cite{peris}.

\appendix
\section{Vertices}
We list below the set of vertices which are needed in sec.2 
and those needed sec.3 which are associated with vector and axial-vector
resonances.
The vertices represent derivatives of the action (times $i$) with respect
to the various sources (with each source multiplied by $i$). 
It is not difficult to derive, in an analogous way, all the other 
relevant vertices.

\subsection{Vertices with pions}
\begin{center}
\begin{texdraw}
\drawdim cm 
\move(0 0)
\move(-8 1)
\textref h:L v:C \htext{$V=-F \partial_\lambda \pi^c(z) F^\lambda(z)$}
\move(0 0)
\textref h:C v:C \htext(1.0 2.1){\qa}
\textref h:L v:C \htext(1.4 2.1){$ q^c_A(z)$}
\linewd 0.04
\setgray 0
\move(1.0 0)
\lvec(1.0 2)
\end{texdraw}
\end{center}
\begin{center}
\begin{texdraw}
\drawdim cm 
\move(0 0)
\move(-8 1)
\textref h:L v:C \htext{$V=iF f^{cat} \pi^t(x) F_\alpha(x)\delta(x-z) $}
\move(0 0)
\textref h:C v:C \htext(1.2 2.1){\qa}
\textref h:C v:C \htext(0.9 2.1){\vmu}
\textref h:R v:C \htext(0.6 2.1){$v_\alpha^a(x)$}
\textref h:L v:C \htext(1.4 2.1){$ q^c_A(z)$}
\linewd 0.04
\setgray 0
\move(1.0 0)
\lvec(1.0 2)
\end{texdraw}
\end{center}
\begin{center}
\begin{texdraw}
\drawdim cm 
\move(0 0)
\move(-8 1)
\textref h:L v:C \htext{$V=iF f^{bct} \pi^t(y) F_\beta(y)\delta(y-z) $}
\move(0 0)
\textref h:C v:C \htext(1.2 2.1){\qv}
\textref h:C v:C \htext(0.9 2.1){\amu}
\textref h:R v:C \htext(0.6 2.1){$a_\beta^b(y)$}
\textref h:L v:C \htext(1.4 2.1){$ q^c_V(z)$}
\linewd 0.04
\setgray 0
\move(1.0 0)
\lvec(1.0 2)
\end{texdraw}
\end{center}
\begin{center}
\begin{texdraw}
\drawdim cm 
\move(0 0)
\move(-8 1)
\textref h:L v:C \htext{$V=iF f^{bat} \pi^t(x)g^{\alpha\beta}\delta(x-y) $}
\move(0 0)
\textref h:C v:C \htext(1.2 2.1){\amu}
\textref h:C v:C \htext(0.9 2.1){\vmu}
\textref h:R v:C \htext(0.6 2.1){$v_\alpha^a(x)$}
\textref h:L v:C \htext(1.4 2.1){$a_\beta ^b(y)$}
\linewd 0.04
\setgray 0
\move(1.0 0)
\lvec(1.0 2)
\end{texdraw}
\end{center}
\begin{center}
\begin{texdraw}
\drawdim cm 
\move(0 0)
\move(-8 1)
\textref h:L v:C \htext{$V=-f^{stc}\partial_\mu\pi^s(z)\pi^t(z)F_\mu(z)$}
\move(0 0)
\textref h:C v:C \htext(1.0 1.0){\qv}
\textref h:L v:C \htext(1.4 1.0){$ q_V^c(z)$}
\linewd 0.04
\setgray 0
\move(1.0 0)
\lvec(1.0 2)
\end{texdraw}
\end{center}
\begin{center}
\begin{texdraw}
\drawdim cm 
\move(0 0)
\move(-8 1)
\textref h:L v:C \htext{$V=-f^{sta}\partial_\alpha\pi^s(x)\pi^t(x)$}
\move(0 0)
\textref h:C v:C \htext(1.0 1.0){\vmu}
\textref h:L v:C \htext(1.4 1.0){$ v_\alpha^a(x)$}
\linewd 0.04
\setgray 0
\move(1.0 0)
\lvec(1.0 2)
\end{texdraw}
\end{center}

\begin{center}
\begin{texdraw}
\drawdim cm 
\move(0 0)
\move(-8 1)
\textref h:L v:C \htext{$V=i\,g^{\alpha\beta}f^{ais}f^{bjs}\pi^i(x)\pi^j(x)\delta(x-y)$}
\move(0 0)
\textref h:C v:C \htext(1.2 1.0){\amu}
\textref h:C v:C \htext(0.9 1.0){\amu}
\textref h:R v:C \htext(0.4 1.0){$ a_\alpha^a(x)$}
\textref h:L v:C \htext(1.4 1.0){$ a_\beta ^b(y)$}
\linewd 0.04
\setgray 0
\move(1.0 0)
\lvec(1.0 2)
\end{texdraw}
\end{center}

\begin{center}
\begin{texdraw}
\drawdim cm 
\move(0 0)
\move(-8 1)
\textref h:L v:C \htext{$V=-i\,g^{\alpha\beta}f^{ais}f^{bjs}\pi^i(x)\pi^j(x)\delta(x-y)$}
\move(0 0)
\textref h:C v:C \htext(1.2 1.0){\vmu}
\textref h:C v:C \htext(0.9 1.0){\vmu}
\textref h:L v:C \htext(1.4 1.0){$ v_\beta^b(y)$}
\textref h:R v:C \htext(0.4 1.0){$ v_\alpha ^a(x)$}
\linewd 0.04
\setgray 0
\move(1.0 0)
\lvec(1.0 2)
\end{texdraw}
\end{center}

\subsection{Resonance vertices with $ v_\mu$ sources: }

These vertices are the same for vector or axial-vector mesons:
\begin{center}
\begin{texdraw}
\drawdim cm 
\move(0 0)
\move(-8 1)
\textref h:L v:C \htext{$V =-f^{a i j} R^i_{\alpha\mu} (x)\partial^\lambda R_{\lambda}^{j \mu}(x)$}
\textref h:C v:C \htext(1.0 1.0){\vmu}
\textref h:L v:C \htext(1.4 1.0){$v^a_\alpha(x)$}
\linewd 0.20
\setgray 0.8
\move(1.0 0)
\lvec(1.0 2)
\end{texdraw}
\end{center}

\begin{center}
\begin{texdraw}
\drawdim cm 
\move(0 0)
\move(-8 1)
\textref h:L v:C
\htext{$V=i f^{ail}\,f^{bjl} R^i_{\alpha\mu}(x)R_{\beta}^{j\mu}(x)\delta(x-y)$}
\move(0 0)
\textref h:C v:C \htext(1.2 1.0){\vmu}
\textref h:C v:C \htext(0.9 1.0){\vmu}
\textref h:L v:C \htext(1.4 1.0){$v^b_\beta(y)$}
\textref h:R v:C \htext(0.4 1.0){$v^a_\alpha(x)$}
\linewd 0.20
\setgray 0.8
\move(1.0 0)
\lvec(1.0 2)
\end{texdraw}
\end{center}

\subsection{Vertices proportional to $F_A$}
\begin{center}
\begin{texdraw}
\drawdim cm 
\move(-8 0)\textref h:L v:R
\htext{$V=-F_A A^c_{\mu \nu}(z)\partial^\mu F^\nu (z)$}
\textref h:C v:C \htext(1.0 2.1){\qa}
\textref h:L v:C \htext(1.4 2.1){$ q^c_A(z)$}
\linewd 0.20
\setgray 0.8
\move(1.0 0)
\lvec(1.0 2)
\end{texdraw}
\end{center}

\begin{center}
\begin{texdraw}
\drawdim cm 
\move(-8 1)\textref h:L v:C
\htext{$
V=iF_A f^{ibc} A^i_{\beta\lambda}(y) F^\lambda(y)\delta(y-z)
$}
\textref h:C v:C \htext(1.2 2.1){\qa}
\textref h:C v:C \htext(0.9 2.1){\vmu}
\textref h:L v:C \htext(1.4 2.1){$ q^c_A(z)$}
\textref h:R v:C \htext(0.4 2.1){$v^b_\beta(y)$}
\linewd 0.20
\setgray 0.8
\move(1.0 0)
\lvec(1.0 2)
\end{texdraw}
\end{center}

\begin{center}
\begin{texdraw}
\drawdim cm 
\move(-8 1)\textref h:L v:C
\htext{$
V=iF_A f^{iad} A^i_{\alpha\lambda}(x) F^\lambda(x)\delta(x)
$}
\textref h:C v:C \htext(1.2 2.1){\qv}
\textref h:C v:C \htext(0.9 2.1){\amu}
\textref h:L v:C \htext(1.4 2.1){$ q^d_V(0)$}
\textref h:R v:C \htext(0.4 2.1){$a^a_\alpha(x)$}
\linewd 0.20
\setgray 0.8
\move(1.0 0)
\lvec(1.0 2)
\end{texdraw}
\end{center}

\begin{center}
\begin{texdraw}
\drawdim cm 
\move(-8 1)\textref h:L v:R \htext{$
V={-F_A\over F} f^{dij} A^i_{\mu\nu}(0)\partial^{\mu}F^\nu(0)\pi^j(0)$}
\textref h:C v:C \htext(0.9 1.0){\qv}
\textref h:R v:C \htext(0.4 1.0){$q^d_V(0)$}
\linewd 0.20
\setgray 0.8
\move(1.0 0)
\lvec(1.0 1)
\linewd 0.04
\setgray 0
\lvec(1.0 2)
\end{texdraw}
\end{center}

\begin{center}
\begin{texdraw}
\drawdim cm 
\move(-8 1)\textref h:L v:R \htext{$
V={F_A\over F} f^{ijl} f^{abl} A^i_{\alpha\beta}(x) \pi^j(x)\delta(x-y)$}
\textref h:C v:C \htext(1.2 1.0){\vmu}
\textref h:C v:C \htext(0.9 1.0){\vmu}
\textref h:L v:C \htext(1.4 1.0){$v^b_\beta(y)$}
\textref h:R v:C \htext(0.4 1.0){$v^a_\alpha(x)$}
\linewd 0.20
\setgray 0.8
\move(1.0 0)
\lvec(1.0 1)
\linewd 0.04
\setgray 0
\lvec(1.0 2)
\end{texdraw}
\end{center}

Note that this vertex is antisymmetric in the Lorentz indices $\alpha\beta$
and will therefore not contribute here. 

\subsection{Vertices proportional to $F_V$}
\begin{center}
\begin{texdraw}
\drawdim cm 
\move(-8 1)\textref h:L v:R
\htext{$V=F_V V^c_{\mu \nu}(z)\partial^\mu F^\nu (z)$}
\textref h:C v:C \htext(1.0 2.1){\qv}
\textref h:L v:C \htext(1.4 2.1){$ q^c_V(z)$}
\linewd 0.20
\setgray 0.8
\move(1.0 0)
\lvec(1.0 2)
\end{texdraw}
\end{center}

\begin{center}
\begin{texdraw}
\drawdim cm 
\move(-8 1)\textref h:L v:C
\htext{$
V=-iF_V f^{iac} V^i_{\alpha\lambda}(x) F^\lambda(x)\delta(x-z)
$}
\textref h:C v:C \htext(1.2 2.1){\qv}
\textref h:C v:C \htext(0.9 2.1){\vmu}
\textref h:L v:C \htext(1.4 2.1){$ q^c_V(z)$}
\textref h:R v:C \htext(0.4 2.1){$v^a_\alpha(x)$}
\linewd 0.20
\setgray 0.8
\move(1.0 0)
\lvec(1.0 2)
\end{texdraw}
\end{center}

\begin{center}
\begin{texdraw}
\drawdim cm 
\move(-8 1)\textref h:L v:C
\htext{$
V=-iF_V f^{iac} V^i_{\alpha\lambda}(x) F^\lambda(x)\delta(x-z)
$}
\textref h:C v:C \htext(1.2 2.1){\qa}
\textref h:C v:C \htext(0.9 2.1){\amu}
\textref h:L v:C \htext(1.4 2.1){$ q^c_A(z)$}
\textref h:R v:C \htext(0.4 2.1){$a^a_\alpha(x)$}
\linewd 0.20
\setgray 0.8
\move(1.0 0)
\lvec(1.0 2)
\end{texdraw}
\end{center}

\begin{center}
\begin{texdraw}
\drawdim cm 
\move(-8 1)\textref h:L v:R \htext{$
V={ F_V\over F}  f^{stc}V^s_{\mu\nu}\partial^\mu F^\nu \pi^t $}
\textref h:C v:C \htext(0.9 1.0){\qa}
\textref h:R v:C \htext(0.4 1.0){$ q^c_A(z)$}
\linewd 0.20
\setgray 0.8
\move(1.0 0)
\lvec(1.0 1)
\linewd 0.04
\setgray 0
\lvec(1.0 2)
\end{texdraw}
\end{center}

\subsection{Vertices proportional to $G_V$}
\begin{center}
\begin{texdraw}
\drawdim cm 
\move(-8 1)\textref h:L v:C
\htext{$
V=2iG_V f^{act} V^t_{\alpha\nu}(x) F^\nu(x)\delta(x-z)
$}
\textref h:C v:C \htext(1.2 2.1){\qa}
\textref h:C v:C \htext(0.9 2.1){\amu}
\textref h:L v:C \htext(1.4 2.1){$ q^c_A(z)$}
\textref h:R v:C \htext(0.4 2.1){$a^a_\alpha(x)$}
\linewd 0.20
\setgray 0.8
\move(1.0 0)
\lvec(1.0 2)
\end{texdraw}
\end{center}

\begin{center}
\begin{texdraw}
\drawdim cm 
\move(-8 1)\textref h:L v:R \htext{$
V= {-2G_V\over F} f^{ast} V^s_{\alpha\mu}(x)\partial^\mu \pi^t(x)
$}
\textref h:C v:C \htext(0.9 1.0){\amu}
\textref h:R v:C \htext(0.4 1.0){$ a^a_\alpha(x)$}
\linewd 0.20
\setgray 0.8
\move(1.0 0)
\lvec(1.0 1)
\linewd 0.04
\setgray 0
\lvec(1.0 2)
\end{texdraw}
\end{center}

\begin{center}
\begin{texdraw}
\drawdim cm 
\move(-8 1)\textref h:L v:R \htext{$
V= {-2G_V\over F} f^{cst} V^s_{\lambda\mu}(z)\partial^\mu \pi^t(z) 
F^\lambda(z)
$}
\textref h:C v:C \htext(0.9 1.0){\qv}
\textref h:R v:C \htext(0.4 1.0){$ q^c_A(z)$}
\linewd 0.20
\setgray 0.8
\move(1.0 0)
\lvec(1.0 1)
\linewd 0.04
\setgray 0
\lvec(1.0 2)
\end{texdraw}
\end{center}

\begin{center}
\begin{texdraw}
\drawdim cm 
\move(-8 1)\textref h:L v:R \htext{$
V= {-i G_V\over F_0^2} f^{stu} \int dv\,
V^s_{\mu\nu}(v)\partial^\mu\pi^t(v)\partial^\nu\pi^u(v)
$}
\linewd 0.20
\setgray 0.8
\move(1.0 0)
\lvec(1.0 1)
\linewd 0.04
\setgray 0
\lvec(2.0 1)
\move(1.0 1)
\lvec(1.0 2)
\end{texdraw}
\end{center}

There are other potentially relevant vertices which are antisymmetric in
$\alpha\beta$ and will thus not contribute for our purposes.

\section{$b_1(1235)$ contributions}
We present here an estimate of the contributions associated with
the $b_1(1235)$ resonance. 
The physics of the $b_1$ meson is not as constrained as that
of the $a_1$ meson by chiral symmetry, so in this section we will occasionally
make some educated guesses. The order of magnitude of the $b_1$ effect
should nevertheless be properly accounted for.
The contributions are to the same correlator
components as those from the $\omega\rho\pi$ couplings. In principle, the
convergence constraints should be applied to the sum. We will apply,however, 
these constraints separately to the $b_1$ for the purpose of obtaining 
an order of magnitude estimate. 

We can write down a set of vertices in exact analogy with
the $a_1\rho\pi$ vertices  
\bea\lbl{lagVBpi}
&&{\cal L}_{b_1\omega\pi}=(G^B_1 g^{\mu_0\mu_1}g^{\mu_2\mu3} +
              G^B_2 g^{\mu_0\mu_2}g^{\mu_3\mu_1} +
              G^B_3 g^{\mu_0\mu_3}g^{\mu_1\mu_2})\trace{
\,\{\nabla_{\mu_0} V_{\mu_1\lambda},B_{\mu_2}^{\phantom{\mu_2}\lambda}\} 
u_{\mu_3}}
\nonumber\\
&&
             +(H^B_1 g^{\mu_0\mu_1}g^{\mu_2\mu_3} +
              H^B_2 g^{\mu_0\mu_2}g^{\mu_3\mu_1} +
              H^B_3 g^{\mu_0\mu_3}g^{\mu_1\mu_2})\trace{
\,\{V_{\mu_1\lambda},B_{\mu_2}^{\phantom{\mu_2}\lambda}\} \nabla_{\mu_0} 
u_{\mu_3}}
\ena
The only difference with the Lagrangian of eq.\rf{lagGH} is that the commutator
is replaced by an anticommutator on account of the fact that the $b_1$ 
meson is odd under charge conjugation. As a consequence, the contribution to
the correlator $\trace{AAQ_VQ_V}_3$ can be deduced simply from the 
results of sec.\ref{sec:quadct}
\bea\lbl{b1exch}
 &&\trace{AAQ_VQ_V}^{b_1\omega\pi}_3=
 {-3F^2_V \over 16\pi^2}\Bigg\{ 
-\left(G^B_1+G^B_2+2G^B_3\right)^2 
\Big[ \logV  +{M^4_B\logBV\over (\mbd-\mvd)^2}
\nonumber\\
&&\phantom{\trace{AAQ_AQ_A}_1= }
+{7\mvd-\mbd\over6(\mvd-\mbd)}\Big]
+4{(G^B_1)^2\mvd\over\mbd}\left(\logV+{5\over3}\right) \Bigg\}\ .
\ena
Imposing a consistency condition on eq.\rf{b1exch}
enables one to express
$\trace{AAQ_VQ_V}^{b_1\omega\pi}_3$ in terms of the 
combination 
\be
\tilde G^B\equiv G^B_1+G^B_2+2G^B_3
\en
and one obtains
\be
\trace{AAQ_VQ_V}^{b_1\omega\pi}_3=
 {3F^2_V \over 16\pi^2}\left(\tilde G^B \right)^2
\Bigg[ {M^4_B\logBV\over (\mbd-\mvd)^2}
+{3\mbd+\mvd\over2(\mbd-\mvd)}  \Bigg]\ .
\en
We must next try to determine the relevant combination of couplings 
$\tilde G^B$ from experiment.
The widths of the decays $b_1^+\to\gamma\pi^+$ and $b_1^+\to\omega\pi^+$
are known
\be\lbl{b1exp}
\Gamma_{b_1^+\to\gamma\pi^+}=227\pm59\ {\rm KeV},\quad
\Gamma_{b_1^+\to\omega\pi^+}=142\pm9\ {\rm MeV}
\en
and their expressions in terms of the resonance Lagrangian coupling constants
read
\bea
&&\Gamma_{b_1^+\to\gamma\pi^+}={\alpha(\mbd-\mpid)\over24 M_B}\left(
{F_V\mbd\over 3\fpi\mvd}\right)^2 \left(\tilde H^B\right)^2
\nonumber\\
&&\Gamma_{b_1^+\to\omega\pi^+}={ p_{cm}(\mbd-M^2_\omega)^2\over 48\pi M^4_B
M^2_\omega\fpid}\Big\{ 2\mbd \mvd \left( \tilde G^B\right)^2
\nonumber\\
&&\phantom{\Gamma_{b_1^+\to\omega\pi^+}={ p_{cm}(\mbd-M^2_\omega)^2\over 48\pi 
M^4_B M^2_\omega\fpid}}
+\left[ (\mbd-\mvd)\tilde H^B+2\mvd\tilde G^B \right]^2  \Big\}\ .
\ena
Here, we have introduced the notation
\be
\tilde H^B\equiv H_1^B+H_2^B+G_2^B+2G_3^B
\en
Using these two pieces of information enables one to determine the combination
of couplings that one needs, but there are several solutions due to the
non linearity of the equations
\be\lbl{sols}
 \tilde H^B=\pm 0.68\ ,
\quad
  \tilde G^B\equiv G_1^B+G_2^B+2G_3^B=\pm 0.24,\ \pm 0.67\ .
\en
We can  eliminate some of the solutions by using one more piece
of experimental information concerning the $D/S$ ratio in the decay
$b_1\to\omega\pi$
\be\lbl{d/sexp}
{f_D\over f_S}=0.29\pm0.04
\en 
The $S$ and the $D$-wave components of the decay amplitude $b_1\to\omega\pi$
(defined as in ref.\cite{isgur}) are easily computed from our Lagrangian
\bea
&&f_S= -{\sqrt{4\pi}\over 3M_BM_\omega F_\pi}\left(\mbd-\mvd\right)
\left[ M_V(M_B+2M_V)\tilde G^B+(\mbd-\mvd)\tilde H^B\right]
\nonumber\\
&&f_D=-{\sqrt{8\pi}\over 6M_BM_\omega F_\pi}\left(\mbd-\mvd\right)
\left[2M_V(M_B-M_V)\tilde G^B+(\mbd-\mvd)\tilde H^B\right]
\ena
Using this result and the experimental one on the $D/S$ ratio reduces the
set of acceptable solutions for the coupling constants to
\be
(\tilde G^B,\tilde H^B)=\pm (0.24, 0.68) 
\en
which give $f_D/f_S=0.28$.

\end{document}